\documentstyle[11pt]{article}
\def\thefootnote{\fnsymbol{footnote}}
\def\su2xsu2{{SU(2)\times SU(2)}}
\def\gL{g_{\mbox{\tiny L}}}
\def\gR{g_{\mbox{\tiny R}}}
\def\LL{\eta}
\def\bea {\begin{eqnarray}}
\def\eea {\end{eqnarray}}
\def\be {\begin{equation}}
\def\ee {\end{equation}}
\def\bitem{\begin{itemize}}
\def\eitem{\end{itemize}}
\def\ie{{\it i.e}}

\def\etal{{\it et. al.}}
\def\del{\partial}
\def\M{{\cal M}}
\def\O{{\cal O}}
\def\T{{\tilde{\cal T}}}
\def\gAnot{{\stackrel{\mbox{\tiny o}}{g}}_{\! A}}
\def\mnot{{\stackrel{\mbox{\tiny o}}{m}}}
\def\Mnot{{\stackrel{\mbox{\tiny o}}{M}}}
\def\Fnot{{\stackrel{\mbox{\tiny o}}{F}}}
\def\Fp {F}
\def\Mp {M}
\def\gA{g_A}
\def\gB {g_A}
\def\pF{p_{\!\mbox{\tiny F}}}
\def\intl{{\frac{\mu^{4-d}}{i} \int \frac{d^d l}{(2\pi)^d}}}
\def\Nbar{{\overline {N}}}
\def\Bbar{{\overline {B}}}

\def\Dslash{{\not {\!\! D}}}
\def\dslash{{\not {\! \partial}}}

\def\vslash{{\not {\! v}}}
\def\pivec{{\vec \pi}}

\def\tauvec{{\vec \tau}}

\def\dmu{{\partial_\mu}}

\def\dmup{{\partial^\mu}}

\def\bfsigma{{\stackrel{\rightarrow}\sigma}}

\def\bfp{{\bf p}}

\def\bfq{{\bf q}}

\def\bfr{{\bf r}}

\def\GpiA{{\Gamma_{\pi A}^{\mu,ab}}}
\def\GpiV{{\Gamma_{\pi V}^{\mu,ab}}}

\def\GVNN{{\Gamma_{VNN}^{\mu,a}}}
\def\GANN{{\Gamma_{ANN}^{\mu,a}}}
\def\GpiNN{{\Gamma_{\pi NN}^a}}
\def\GVpipi{{\Gamma_{V\pi\pi}^{\mu,abc}}}

\def\dab{{\delta_{ab}}}
\def\eabc{{\epsilon_{abc}}}

\def\Pp{{ \, P_+ \,}}
\def\Pa{{ \, P_- \,}}
\def\Bp{{B^{(+)}}}
\def\Ba{{B^{(-)}}}
\def\Bbarp{{{\overline {B}}^{(+)}}}

\def\prl {Phys. Rev. Lett.}
\def\pl {Phys. Lett.}

\def\np {Nucl. Phys.}
\def\ct{{\mbox{\tiny CT}}}
\newcommand{\e}{{\mbox{e}}}

\begin{document}

\begin{titlepage}
\begin{center}
\begin{flushright}
SNUTP 92-97
\end{flushright}
\hfill\today
\vskip 1.0cm
{\large\bf Chiral Dynamics and Heavy-Fermion Formalism}
\vskip 0.2cm
{\large\bf in Nuclei: I. Exchange Axial Currents}
\vskip 2cm
{\large Tae-Sun Park}\\
{\it Department of Physics and Center for Theoretical Physics}\\
{\it Seoul National University}\\
{\it Seoul 151-742, Korea}
\vskip 0.2cm
{and}
\vskip 0.2cm
{\large Dong-Pil Min\footnote{Permanent address: Department of Physics, Seoul
National University, Seoul 151-742, Korea.} and Mannque Rho}\\
{\it Service de Physique Th\'{e}orique, CEA Saclay}\\
{\it 91191 Gif-sur-Yvette, France}
\vskip 2cm
{\bf ABSTRACT}
\begin{quotation}
Chiral perturbation theory in heavy-fermion formalism is developed for
meson-exchange currents in nuclei and applied to nuclear axial-charge
transitions. Calculation is performed to the next-to-leading order
in chiral expansion which involves graphs up to one loop. The result
turns out to be very simple. The previously
conjectured notion of ``chiral filter mechanism" in the time component
of the nuclear axial current and the space component of the nuclear
electromagnetic current is verified to that order. As a consequence,
the phenomenologically observed soft-pion dominance in the nuclear process
is given a simple interpretation in terms of chiral symmetry in nuclei.
In this paper we focus on the axial current, relegating the electromagnetic
current which can be treated in a similar way
to a separate paper. We discuss the implication of our result
on the enhanced axial-charge transitions observed in heavy nuclei and clarify
the relationship between the phenomenological meson-exchange description
and the chiral Lagrangian description.
\end{quotation}
\end{center}
\end{titlepage}
\renewcommand{\thefootnote}{\#\arabic{footnote}}
\setcounter{footnote}{0}
\section{Introduction}
\indent \indent By now
there exist a large number of unambiguous experimental evidences \cite{fm} for
meson-exchange currents in nuclear responses to electroweak probes. We also
have available a rather satisfactory and successful theory to describe
the large bulk of experimental observations \cite{review}. While inherently
phenomenological in character, the approaches taken so far
to describe meson-exchange
currents have been commensurate with the ingredients that account
for our progressive understanding of nuclear forces
and to the extent that nucleon-nucleon interactions are now fairly accurately
understood, one can have a great deal of confidence in
the theoretical tool with which the effect of exchange currents
is calculated. There remains however the fundamental question as to how
our phenomenological understanding of nuclear forces and associated
meson currents can be linked
to the fundamental theory of strong interactions, QCD.

In this paper we make a first step towards answering this question
by applying chiral perturbation theory (ChPT) to nuclear electroweak
processes. To start with, we assume that
at low energies dominated by infrared properties of QCD, the most important
aspect of QCD is the spontaneously
broken chiral symmetry and hence that in nuclear dynamics, it
is chiral symmetry that plays a predominant role \cite{rb}. The important role
of chiral symmetry in nuclear physics
was recognized early on by Chemtob and Rho for exchange currents \cite{chemrho}
but this issue was recently given a stronger impetus and a more modern meaning
by Weinberg in connection with nuclear
forces \cite{wein90} and by Rho \cite{mr91} in connection with what is known
``nuclear chiral filter phenomenon" (for definition, both intuitive and
more rigorous, see later).

The key question addressed here is this:
To what extent can nuclear processes be described by QCD or equivalently
at low energies by
chiral perturbation theory? Weinberg approaches this issue by studying
nuclear many-body forces. We propose here to do the same by looking at
nuclear response functions responding
to slowly varying electroweak fields
\cite{pmr92}. We suggest that chiral perturbation theory can be
made -- under certain conditions specified below -- considerably more
powerful and predictive for response functions than for nuclear forces.
In calculating nuclear forces to loop orders
in chiral perturbation theory, one encounters a plethora
of counter terms to renormalize the theory, most
of which are not accessible by experiments \cite{ordonez}; furthermore there
are contact four-fermion interactions in the Lagrangian -- most of which
are again unknown parameters -- that have to be carefully
examined and treated. In principle, this may be feasible, perhaps
with the help of lattice QCD calculations
but in practice it may not be possible to make clear and
useful predictions because of
many uncontrolled parameters. While the tree order chiral theory justifies
{\it \`{a} posteriori} the current nuclear physics practice of using two-body
static forces \cite{wein90},
it appears that chiral symmetry will be unable
to make any truly significant statement on the structure of nuclear forces for
sometime to come. A major new development will be required before one can
make a prediction that goes beyond the accuracy of the phenomenological
approach which has been strengthened by the wealth of experimental data.
On the contrary, as we will show
in this paper, when the formalism is applied to nuclear response
functions, in particular, to exchange currents, it can make a highly
nontrivial and potent prediction. This is because nuclear short-range
correlations generated by nuclear interactions at short distance
-- which while poorly understood of their mechanism, are nonetheless
operative in nuclear medium -- screen
all the contact interactions, both intrinsic and induced
and consequently {\it all} of the four-fermion (and higher)
counter term contributions, effectively ``filtering" off the ill-understood
short-range operators: Given phenomenological information on nuclear
wave functions at short distance, the short-range suppression helps
in simplifying nuclear response functions.
In addition, in certain kinematic conditions, higher
order chiral corrections are found to be {\it naturally} suppressed. The
suppression of the many-fermion counter term contributions at the one-loop
order that we are studying is, as will be stated more precisely later,
a consequence of the fact that such terms occurring
at high orders in the chiral expansion reflect the degrees of freedom
that enter directly neither in nuclear forces nor in nuclear currents
at the chiral order considered.
The combination
of these two phenomena lead to the ``chiral filtering" proposed
previously \cite{kdr}. In this paper, we will establish this chiral filtering
to one-loop order in chiral perturbation theory. This will provide,
in our opinion,
the very first compelling explanation of the pion-exchange
dominance observed in axial charge transitions (considered in this paper)
as well as
in radiative np capture or in threshold electrodisintegration of the
deuteron.

As stressed by Weinberg, chiral perturbation theory is useful in
nuclear physics {\it
only for ``irreducible" diagrams} that are by choice free of infrared
divergences. This means that both in nuclear forces and in exchange currents,
reducible graphs are to be taken care of by a Schr\"{o}dinger equation
or its relativistic generalization
with the irreducible graphs entering
as potentials. This also implies that in calculating exchange currents
in ChPT, we are to use the wave functions so generated to calculate matrix
elements
to obtain physical amplitudes. This is of course the standard practice in
the theory of meson-exchange currents but it is also in this sense that
ChPT is predictive in nuclei. Clearly this precludes what one might call
``fully consistent chiral perturbation theory" where nuclear forces,
nuclear currents and wave functions are all calculated to the same order
of chiral perturbation expansion.  Such a calculation even if feasible
is likely to make no sense. A little thought would persuade the reader
that it is a futile exercise.

We will here focus on
the irreducible diagrams contributing to exchange currents. We will
calculate next-to-leading order terms in the chiral counting
involving one-loop graphs. In doing this, we will employ
the recently developed heavy fermion formalism (HFF) \cite{jm91}.
The standard ChPT \cite{gl} arranges terms in power of $(\del/\Lambda_\chi)$
and/or of $(m_\pi/\Lambda_\chi)$ where $\del$ is four-derivative
acting on the Goldstone boson ({\it viz}, pion) field, $m_\pi$ the
pion mass ($\approx 140$ MeV) and $\Lambda_\chi\approx 1$ GeV, the
chiral expansion scale. It has been established that this expansion
works well at low energies for such processes as $\pi\pi$ scattering.
However the situation is different when baryons are involved. The
dynamically generated masses of the baryons are of $O(\Lambda_\chi)$
and hence when the baryon field is acted upon by time-derivative,
it gives an $O(1)$ term. Therefore a straightforward derivative
expansion fails. (Incidentally this is also the reason why
a chiral Lagrangian describing pion interactions well with low-order
derivative terms does not necessarily describe well skyrmion properties.)
The HFF circumvents this difficulty in rearranging the derivative expansion.
Indeed the principal virtue of the HFF is that
it provides a consistent chiral expansion
in $Q/\Lambda_\chi$ where $Q$ is four-derivative on pion field or pion mass
or space derivative on baryon field; it avoids time derivative on baryon
field which is of order $\Lambda_\chi$ which is not small. The standard
ChPT involving baryons \cite{gss} can in principle be arranged to give
a similar expansion. However it requires a laborious reshuffling of terms
avoided in the HFF.
The distinct advantage of the HFF is that the multitude of diagrams
that appear in such calculations as ours in the standard ChPT involving baryons
get reduced to a handful of manageable terms, thus alleviating markedly
the labor involved. We will see that there is an enormous simplification
in the number of terms and in their expressions.
The potential disadvantage might be that the HFF is not fully
justified for the mass corresponding to that of the nucleon and
hence higher order ``$1/m$" corrections may have to be systematically included.
We will examine the class of approximations we make in the calculation
by looking at the next order terms. It turns out that the leading
``$1/m$" correction is absent in our calculation. We shall discuss this
matter in the concluding section.

While the procedure is practically the same, the resulting expression for
electromagnetic (EM) current is somewhat more involved. We will therefore
not treat it here although we shall give a general treatment of
the theory applicable to both axial and EM currents. The detailed
analysis on the EM currents, together with an application to threshold
np capture, will be reported in a separate paper \cite{pmr2}. Both currents
are intricately connected even at low energy through current algebras and
we will need some vertices involving the EM current.

It is perhaps obvious but we should stress that
for both vector and axial-vector currents, relevant symmetries ({\ie},
conserved vector current and partially conserved axial-vector current)
are preserved to the chiral order considered since both nuclear
forces and currents are treated on the same footing with the same effective
Lagrangian. More on this point later.

This paper is organized as follows. In Section 2, we state our basic assumption
in applying ChPT to nuclear dynamics. In Section 3, we describe
the effective chiral Lagrangian with which we develop heavy-fermion formalism
including ``$1/m$" corrections. We also define the relevant kinematics we will
consider. The chiral counting rules are given in Section 4. In Section 5,
the renormalization of n-point vertices that enter in the calculation
is detailed. For the sake of making this paper as self-contained as
possible and to define notations,
we also list the renormalized quantities for the pion and the nucleon
following from the Lagrangian. Readers familiar with renormalization
of heavy-fermion chiral Lagrangian could proceed directly to subsection
5.4. Two-body exchange currents are calculated in Section 6. Both
momentum-space and coordinate-space formulas are given. Numerical analyses
are described in Section 7. In Section 8, we explain why there are {\it
no other graphs} that can contribute to the same order and point out
in what circumstances they can show up in physical observables.
Concluding remarks including those on
the observed enhanced axial-charge transitions in heavy nuclei are
made in Section 9.
The Appendices A-I list all the formulas needed in the calculation.

This paper is written in as a self-contained way as possible so as to
be readable by those who are not familiar with the recent development
in the field. Some of the material are quite standard and
readily available in the literature. Most of them however serve as a
check of our calculation.

\section{Strategies in Nuclear Physics}
\indent

We wish to calculate operators effective in nuclei for transitions
induced by the vector and axial vector currents of electroweak interactions,
denoted respectively by $V_\mu$ and $A_\mu$ associated with the electroweak
fields ${\cal V}_\mu$ and ${\cal A}_\mu$. In principle there will be
$n$-body currents for $N\geq n >1$ in $N$-body systems.
Here we will focus only on
one- and two-body currents, ignoring those with $n>2$. The reasons for
so doing are given in the literature \cite{review,chemrho} but we will later
show that $n$-body currents for $n>2$ are suppressed to the order considered
for long wavelength probes.

The diagrams we wish to calculate are generically given by Fig. 1. They
correspond to the standard definition of
single-particle and two-particle exchange currents entering in the
description of nuclear response functions to the external electroweak fields.
These have been calculated before in terms of phenomenological Lagrangians.
Here we wish to do so using chiral perturbation theory (ChPT), starting with
a chiral Lagrangian that is supposed to model QCD at low energies.
Following the
chiral counting rule we will derive later, we will restrict our
consideration to one-loop order, which corresponds to going to the
next order in the chiral counting to the leading soft-pion limit.
Although one-loop calculations have been done before for nucleon properties
\cite{bkmp} and for infinite nuclear matter \cite{pw91}, they have up to date
not been performed in finite nuclear processes. We believe this work
constitutes the first attempt to implement consistently chiral symmetry in
nuclear processes.

In dealing with divergences encountered in calculating loop graphs,
in particular the loops involving two-pion exchange, we will need
a certain prescription for handling operators that are short-ranged in
coordinate space. This prescription does not follow from chiral symmetry
alone and will have to be justified on a more general ground.
Specifically, we argue that consistency with ChPT demands that zero-range
interactions be ``killed" by nuclear short-range correlations:
\bitem
\item
Firstly, the zero-range operators that come from finite counter terms
appearing in four-fermion interactions figure neither directly nor
indirectly -- but importantly -- in the successful phenomenological
nucleon-nucleon potentials and hence must represent the degrees of freedom
unimportant for the length scale involved.
In fact, one can show (see Appendix I) that the counter terms we need to
introduce (denoted $\kappa_4^{(1,2)}$ later)
{\it cannot} arise, unlike in the better understood $\pi$-$\pi$ scattering
\cite{ecker}, from an approximation of taking an infinite mass limit of
the strong interaction resonances such as the vector mesons $\rho$, $\omega$
etc. which have a scale comparable to the chiral scale and play an important
role in boson-exchange potential models.
\item Secondly, ChPT by its intrinsic limitation cannot possibly provide
a nuclear force that can
account for the interactions shorter-ranged than two pion or one vector-meson
range at most. Thus the truly short-range interaction known to be present
in the nucleon-nucleon interaction must involve elements that are not
calculable by means of finite-order
chiral expansion even if such an expansion existed.
Thus it would be inconsistent to put a part of such interactions into the
currents in the context of ChPT without a similar account in the
nuclear force. It is known that even to one-loop order, the number of
counter terms is so large in the calculation of nuclear forces that
it is highly unlikely that one can make a meaningful prediction based strictly
on low-order chiral perturbation expansion \cite{ordonez}. As suggested in
Ref.\cite{wein92}, one should implement ChPT calculations
with phenomenological informations whenever available.
\item  Applied to the ``irreducible diagrams" that enter in the
definition of exchange currents, ChPT screens out the short-range
part of the interaction which originates from dynamics of possibly
non-chiral origin. When the matrix elements of the operators arising from
the irreducible graphs are calculated with wave functions suitably computed
in the presence of two-nucleon potentials,
the short-range correlation built into nuclear wavefunctions
must therefore suppress strongly
interactions that occur at an internuclear distance
$\leq 0.6$ fm, automatically ``killing" the $\delta$ function
interactions associated with finite counter terms. This fact will be kept in
mind when we derive two-body operators in coordinate space.
\eitem
\noindent
We will present, at several places in the paper, arguments
to justify the above procedure which purports to establish that
{\it the only unknown
parameters in the theory must be (a) negligible in magnitude  and (b) further
suppressed by nuclear correlations when embedded in nuclear medium.}

There is nothing very much new in our calculation of the one-body operators
except for its consistency with chiral invariance. As for the two-body exchange
currents, our results are new.
There are two graphs to consider: One-pion exchange (Fig. 2a)
and two-pion exchange (Fig. 2b). Both involve one-loop order graphs.
Note that we are to calculate only ``irreducible graphs."

\section{Effective Chiral Lagrangian}
\indent \indent We begin with the effective chiral Lagrangian that consists of
pions and nucleons involving lowest derivative terms\cite{gss} relegating
the role of other degrees of freedom such as vector mesons and nucleon
resonances $\Delta$ to a later publication\footnote{While the vector mesons
and the nucleon resonances (in particular, the $\Delta$) play an important
role in nuclear phenomenology -- and they can be easily implemented in ChPT
at least in low orders, they are unimportant for the process we discuss in this
paper. It is not difficult to see which processes require such degrees of
freedom but we will not pursue this matter, for a treatment of such processes
goes beyond the framework of ChPT.},
\bea
{\cal L}_0 &=&
\Nbar\left[ i\gamma^\mu (\dmu + \Gamma_\mu) - m
+ i \gA \gamma^\mu\gamma_5 \Delta_\mu\right] N
 - \frac12 C_a \left(\Nbar \Gamma_a N\right)^2
\nonumber \\
&&+\, \frac{F^2}{4} {\rm Tr}\left(\nabla_\mu \Sigma^\dagger
\nabla^\mu \Sigma\right) + \frac12 M^2 F^2 {\rm Tr}(\Sigma)
+ \cdots + {\cal L}_\ct,\label{chiralag}
\eea
where $m\simeq 939$MeV is the nucleon mass, $\gA\simeq 1.25$
is the axial coupling constant and $F \simeq 93$ MeV is the pion decay
constant.
The ellipsis stands for higher derivative and/or symmetry-breaking
terms which will given later as needed. We have written the Lagrangian
with the renormalized parameters $m$, $g_A$, $F$ and $M$ with suitable
counter terms ${\cal L}_\ct$ to be specified later.
Under the chiral SU(2)$\times$SU(2) transformation\footnote{We are using a
slightly unconventional notation of Ref. \cite{gss} which we will
follow in this paper. This facilitates checking our results on single-nucleon
properties against those derived in \cite{gss} using standard (relativistic)
chiral
perturbation expansion. The more familiar transformation of the chiral field
used in the literature is gotten by replacing $\Sigma$ by $\Sigma^\dagger$.
We are also working with the exponentiated (Sugawara) form of
chiral Lagrangian instead of Weinberg's \cite{wein90} used previously.
They are of course equivalent. For the rest we will follow the Bjorken-Drell
metric and convention.},
the chiral field $\Sigma= {\rm exp}(i\frac{\tauvec\cdot \pivec}{F})$
transforms as $\Sigma\rightarrow \gR \Sigma \gL^\dagger$
($\gR,\gL \in$ SU(2))
and the covariant derivative of the chiral field transforms as $\Sigma$
does,
\bea
\nabla_\mu \Sigma &=& \dmu \Sigma - i ({\cal V}_\mu+{\cal A}_\mu) \Sigma
+ i \Sigma ({\cal V}_\mu - {\cal A}_\mu)
\nonumber\\
&\rightarrow & \gR\, \nabla_\mu \Sigma \, \gL^\dagger
\eea
where the external gauge fields
${\cal V}_\mu= {\vec {\cal V}}_\mu \cdot \frac{\tauvec}{2}$ and
${\cal A}_\mu= {\vec {\cal A}}_\mu\cdot \frac{\tauvec}{2}$
transform locally
\bea
{\cal V}_\mu+{\cal A}_\mu \rightarrow
{\cal V}_\mu' + {\cal A}_\mu' = \gR ({\cal V}_\mu + {\cal A}_\mu)\gR^\dagger
-i\dmu \gR \cdot \gR^\dagger,
\nonumber \\
{\cal V}_\mu-{\cal A}_\mu \rightarrow
{\cal V}_\mu' - {\cal A}_\mu' = \gL ({\cal V}_\mu - {\cal A}_\mu)\gL^\dagger
-i\dmu \gL \cdot \gL^\dagger.
\nonumber \label{vaLR}\eea
In our work, only the electroweak ($SU(2)\times U(1)$) external fields
will be considered. The Lagrangian of course has global $\su2xsu2$ invariance
in the absence of the pion mass term.
Non-linear realization of chiral symmetry is expressed in terms of
$\xi = \sqrt{\Sigma} = {\rm exp}(i\frac{{\vec \tau}\cdot {\vec \pi}}{2 F})$
and $U = U (\xi,\gL,\gR)$ defined with $\xi$
$$\xi \rightarrow \gR \xi U^\dagger = U \xi \gL^\dagger.$$
Now nucleon field $N$ transforms as $N\rightarrow U N$, and covariant
derivatives of nucleon field and chiral field transform as nucleon field does,
$D_\mu N \rightarrow U D_\mu N$ and
$\Delta_\mu \rightarrow U \Delta_\mu U^\dagger$ where\footnote{
We have defined two covariant derivatives involving chiral fields,
$\nabla_\mu \Sigma$ and $\Delta_\mu$. We can express one in terms of the
other, but it is convenient as done frequently in the literature to use
$\nabla_\mu \Sigma$ for the meson sector and $\Delta_\mu$ for the
meson-nucleon sector.}
\bea
D_\mu N &=& (\dmu + \Gamma_\mu) N ,
\nonumber \\
\Gamma_\mu &=& \frac{1}{2} \left[\xi^\dagger,\dmu \xi\right]
-\frac{i}{2}\xi^\dagger ({\cal V}_\mu+{\cal A}_\mu) \xi - \frac{i}{2}
\xi ({\cal V}_\mu - {\cal A}_\mu)\xi^\dagger,
\nonumber\\
\Delta_\mu &=& \frac12 \xi^\dagger \left(\nabla_\mu \Sigma\right) \xi^\dagger
= \frac12 \left\{ \xi^\dagger , \dmu \xi\right\}
- \frac{i}{2} \xi^\dagger ({\cal V}_\mu+{\cal A}_\mu) \xi
+ \frac{i}{2} \xi ({\cal V}_\mu - {\cal A}_\mu)\xi^\dagger.
\label{deltamu}\eea
The $U$ can be expressed as a complicated local function of $\xi$,
$\gL$ and $\gR$.
The explicit form of $U$ is not needed for our discussion.

Note that we have included the four-fermion non-derivative contact term
studied recently by Weinberg\cite{wein90}. We will ignore possible four-fermion
contact terms involving derivatives (except for counter terms encountered
later) and quark mass terms since they are not relevant
to the chiral order (in the sense defined precisely later) that we are working
with. The explicit chiral symmetry breaking is included minimally in the
form of the pion mass term.
Higher order symmetry breaking terms do not
play a role in our calculation.
\subsection{Heavy-fermion formalism}
\indent\indent
For completeness -- and to define our notations, we sketch here
the basic element of the heavy-fermion formalism
(HFF)\cite{georgi} applied to nuclear systems as developed by Jenkins
and Manohar\cite{jm91} wherein the nucleon is treated as a heavy fermion.
As stressed in Introduction, the relativistic formulation of ChPT works
well when only mesons are involved but it does not work when baryons are
involved since while space derivatives on baryons fields can be arranged
to appear on the
same footing as four-derivatives on pion fields, the time derivative on baryon
fields picks up a term of order of the chiral symmetry breaking scale and hence
cannot be used in the chiral counting. This problem is avoided in the
HFF. To set up the HFF, the fermion momentum is written as
\be
p^\mu = m v^\mu + k^\mu
\ee
where $v^\mu$ is the $4-$velocity with $v^2=1$, and $k^\mu$ is the small
residual momentum. (In the practical calculation that follows, we will choose
the heavy-fermion rest frame $v^\mu =(1, \vec{0})$.)
We define heavy fermion field $B_v(x)$ for a given
four-velocity $v^\mu$,
by\footnote {Another familiar field redefinition is
$B_v(x) = \e^{i m \gamma\cdot v\, v\cdot x} N(x)$,
with $v^2=1$. This definition gives exactly the same physics to the
lowest order in $\frac{1}{m}$ expansion.}
\be
B_v(x) = \e^{i m \, v\cdot x} N(x).
\ee
The field $B_v$ is divided into two parts which are eigenstates of $\vslash$,
\be
B_v= B_v^{(+)} + B_v^{(-)} \equiv
\frac{1+\vslash}{2} B_v + \frac{1-\vslash}{2} B_v
\equiv \Pp B_v + \Pa B_v.
\ee
As defined, $B_v^{(+)}(B_v^{(-)})$ can be identified as positive
(negative) energy solution. As will be justified in the following
subsection, the negative energy solution is suppressed for large baryon mass
and its contribution can subsequently be incorporated as higher-order
corrections in the inverse mass expansion. Thus to the leading order,
the fermion loops can be ignored.
With the neglect of the negative energy solutions, we have
a useful relation for gamma matrices sandwiched  between spinors
which holds for any $\Gamma$,
\be
\Bbar_v \Gamma B_v = \Bbar_v \vslash \Gamma B_v
= \Bbar_v \Gamma \vslash B_v = \Bbar_v \frac{1}{2}
\{ \Gamma , \vslash \} B_v\,.
\ee
It follows from this identity that
\be
\Bbar_v \gamma_5 B_v = 0,\ \
\Bbar_v \gamma^\mu B_v = v^\mu \Bbar_v B_v.
\ee
Let us define spin operators $S_v^\mu$ by
\be
\Bbar_v \gamma_5 \gamma^\mu B_v = -2 \Bbar_v S_v^\mu B_v
\ee
or explicitly
\be
S_v^\mu = \frac14 \gamma_5 \left[ \vslash, \gamma^\mu \right] \,.
\ee
The spin operators have the following identities,
\bea
\left\{ S_v^\mu, S_v^\nu\right\} &=& \frac12 (v^\mu v^\nu - g^{\mu\nu})\, ,\\
\left[ S_v^\mu, S_v^\nu\right] &=& i \epsilon^{\mu\nu\alpha\beta}
                          v_\alpha S_\beta\,\ \ \
\mbox{with} \ \ \ \epsilon_{0123}=1\,.
\eea
{}From the anti-commutation rule, we have
\bea
S_v \cdot S_v &=& \frac14 (1- d) =
-\frac{3- 2\epsilon}{4} \simeq - \frac34\, , \\
S_v^\alpha S_v^\mu S_{v\alpha} &=& \frac14 (d-3) S_v^\mu
\simeq \frac14 S_v^\mu, \\
\left( q \cdot S_v\right)^2 &=&
\frac14 \left[ (q\cdot v)^2 - q^2\right]
\eea
where $d$ is the dimension of the space-time, $d= g^\mu_\mu$ and
we have defined $\epsilon=(4-d)/2$.
Between spinors, we have the approximate relations
\bea
\Bbar_v \, S_v^\mu \, B_v &\simeq& \left(
\frac12 {\vec \sigma}\cdot{\vec v},\,\, \frac12 {\vec \sigma}\right),
\\
\Bbar_v  \left[ S_v^0, {\vec S_v}\right] B_v &\simeq&
-\frac{i}{2} {\vec v} \times {\vec \sigma}
\eea
with ${\vec \sigma}$ the usual Pauli spin matrices.  We see that $S_v^0$ and
$\left[ S_v^0, {\vec S_v}\right]$ are suppressed by a factor of
${\vec v}= {\cal O}\left(\frac{Q}{m}\right)$ where $Q$ is the characteristic
small momentum scale for processes with small three-velocity.
Since
\bea
\left[S^\mu_v , S^\nu_v\right] &=& \frac{i}{4} \left( \sigma^{\mu\nu}
+ \vslash \sigma^{\mu\nu}\vslash\right)
\nonumber \\
&=& \frac{i}{2} \sigma^{\mu\nu} + \left(v^\mu S^\nu_v - v^\nu S^\mu_v\right)
\gamma_5
\nonumber \eea
where $\sigma^{\mu\nu}= \frac{i}{2}\left[\gamma^\mu ,\gamma^\nu\right]$,
we also have
\bea
\Bbar_v \left(\frac{i}{2} \sigma^{\mu\nu} \right)B_v
&=& \Bbar_v \left[ S^\mu_v, S^\nu_v\right] B_v
\\
\Bbar_v \left(\sigma^{\mu\nu} \gamma_5\right)B_v
&=& 2i \Bbar_v \left( v^\mu S^\nu_v- v^\nu S^\mu\right) B_v .
\eea

We are now in position to write down the chiral Lagrangian (\ref{chiralag})
in HFF. The nucleon part of the Lagrangian becomes
\be
\Nbar( i \dslash - m) N = \Bbar_v iv\cdot \partial B_v
\ee
and the corresponding nucleon propagator $S(mv+k)$ is\footnote{
Although we do not actually encounter it in our calculation,
it might be worthwhile to point out one technical subtlety.
The infinitesimal $i 0^+$ is inserted to define the singularity structure
of the propagator.  When we encounter a $d-$dimensional loop integral
we first perform the Wick-rotation to put it in the Euclidean
metric. In doing this, we assume that the first and third quadrants
(in the plane of real  $l^0$ vs. imaginary $l^0$)
contain no poles. If we can take the flow direction of the loop-momentum
to be the direction of the fermion momentum, there is no problem.
However for some graphs it is impossible to do this. For instance
consider a two-nucleon box diagram. In this case,  we have one fermion
line in which the loop-momentum flow direction is opposite to that of the
fermion arrow. In this case, the fermion propagator is of the form
$$\frac{1}{v\cdot k -i 0^+} = \frac{1}{v\cdot k+i0^+}
+2i \pi \delta(v \cdot k).$$ }
\be
i S(mv+k) = \frac{i}{v\cdot k + i 0^+}\, .
\ee
Our chiral Lagrangian (\ref{chiralag}) expressed
in terms of the heavy-fermion field to leading ({\ie}, zeroth)
order in $\frac{1}{m}$ takes the form
\bea
{\cal L}_0 &=& \Bbar_v\left[ i v^\mu (\dmu + \Gamma_\mu)
+ 2 i \gA S_v^\mu \Delta_\mu\right] B_v
- \frac12 C_a \left(\Bbar_v \Gamma_a B_v\right)^2
\nonumber \\
&&+\, \frac{F^2}{4} {\rm Tr}\left(\nabla_\mu \Sigma^\dagger
\nabla^\mu \Sigma\right) + \frac12 M^2 F^2 {\rm Tr}(\Sigma) .
\label{chiralag2} \eea
In practical calculations, the chiral field $\Sigma$ or $\xi$ is expanded
in power of the pion field. The explicit form resulting from such expansion
as well as the vector and axial-vector currents calculated via Noether's
theorem are given in Appendix A.
\subsection{$\frac{1}{m}$ corrections}
\indent \indent
As mentioned, the HFF is based on simultaneous expansion
in the chiral parameter and in ``$1/m$".
We have so far considered leading-order terms in $1/m$, namely, $O((1/m)^0)$.
We now discuss $\frac{1}{m}$ corrections following closely the discussion
of Grinstein \cite{grin}. We choose to do this
in a perhaps more general way than needed for our purpose.
Consider the following Lagrangian
\bea
{\cal L} &=& \Nbar \left[i \Dslash -m + \gamma^\mu \gamma_5 A_\mu\right] N
- \frac12 C_a (\Nbar \Gamma_a N)^2
\nonumber \\
&=& \Bbar \left[i \Dslash -m(1-\vslash) + \gamma^\mu \gamma_5 A_\mu\right] B
- \frac12 C_a (\Bbar \Gamma_a B)^2
\nonumber \eea
where we have included Weinberg's four-fermion contact term with $\Gamma_a$
an arbitrary hermitian operator which we assume to contain no derivatives.
We have also introduced an arbitrary ``axial" field $A_\mu$ which we take to be
hermitian and free of gamma matrices. (Here and in what follows, we shall
omit the subscript $v$ in $B_v$, $\Bbar_v$ and $S_v^\mu$.)
The equation of motion satisfied by $B$ is
\be
\left[G - m(1-\vslash)\right]B=0
\label{eq}\ee
with
\be
G\equiv g - C_a \Gamma_a (\Bbar \Gamma_a B),\ \ \
g\equiv i\Dslash + \gamma^\mu \gamma_5 A_\mu.\label{defg}
\ee
Multiplying (\ref{eq}) on the left by $\Pa$, we obtain
$$ \Pa G\, B - 2m \Ba = 0$$
which leads to
\be \Ba = \frac{1}{2m} \Pa G\, B =
\frac{1}{2m} \Pa G\, \Bp + {\cal O}\left(\frac{1}{m^2}\right).
\label{Bminus}\ee
Now multiplying $\Pp$ to (\ref{eq}), we get
$$\Pp  G\,B =0$$
which gives
\be
\Pp \left[ G + \frac{1}{2m} G \Pa G\right] \Bp
= {\cal O}\left(\frac{1}{m^2}\right).
\label{Bplus}\ee
Given this, it is now a simple matter to write down the Lagrangian that gives
rise to the equation of motion to the desired order
\footnote{We remind the reader that one should not insert the
solution of $\Ba$ into the original Lagrangian, since in Lagrangian
approach, what is important is the form, not the value. However in
Hamiltonian approach, the insertion of the solution for $\Ba$ into
the original Hamiltonian is allowed.}.
The result correct to $O(1/m)$ is
\bea
{\cal L} &=& \Bbar \left[ g + \frac{1}{2m} g \Pa  g\right] B
\nonumber \\
&& - \frac12 C_a \left\{\Bbar \left[\Gamma_a + \frac{1}{2m} \left(
\Gamma_a \Pa g + g \Pa\Gamma_a\right)\right]B\right\}^2
\nonumber \\
&&+ \frac{1}{2m} C_a C_b \left(\Bbar \Gamma_a B\right)
\left(\Bbar \Gamma_a \Pa \Gamma_b B\right)
\left(\Bbar \Gamma_b B\right)
\eea
with $g$ defined in (\ref{defg}). To put this into a more standard form, we use
the identities
\bea
\Pp g \Pp &=& \Pp \frac12 \left\{\vslash , g\right\} \Pp ,
\nonumber \\
\Pp g\Pa g' \Pp &=& \Pp \left\{\Pa , g\right\} g' \Pp =
\Pp g \left\{\Pa , g' \right\} \Pp ,
\nonumber \\
\Pp g \Pa g \Pp &=& \Pp \left[\frac12 \left\{\vslash , g^2\right\}
-  \frac14\left\{g, g\,\left\{\vslash , g\right\} \right\}\right] \Pp .
\eea
One can show from these identities  that
\be \left( \Bbarp \, \Gamma_a \, \Bp\right) \otimes
\Bbarp \, \Gamma_a \, \Pa = 0
\ee
for any $\Gamma_a = \left\{
1,\ \gamma_5,\ \gamma_\mu, \ \gamma_\mu\gamma_5,\ \sigma_{\mu\nu}\right\}$.
This allows us to simplify the Lagrangian further to the form
\bea
{\cal L} &=& \Bbar \left( i v \cdot D + 2 S\cdot A \right) B
-\frac12 C_a \left(\Bbar \Gamma_a B\right)^2
+ \frac{1}{2m}  \Bbar \left( - D^2 + (v\cdot D)^2 \right. \nonumber\\
&& \left.  + \left[S^\mu ,
S^\nu \right] \left[D_\mu,D_\nu\right]- (v\cdot A)^2
-2i \left\{v\cdot A, S\cdot D\right\}\right) B,
\eea
for general $\Gamma_a$ allowed by symmetries.
In our case, $A_\mu = i \gA \Delta_\mu=i \gA \frac 12 \xi^\dagger (\nabla_\mu
\Sigma)\xi^\dagger$, so our
$\frac{1}{m}$ term Lagrangian is
\bea \delta {\cal L} &=&
\frac{1}{2m}  \Bbar \left( - D^2 + (v\cdot D)^2 + \left[S^\mu ,
S^\nu \right] \left[D_\mu,D_\nu\right]+ \gA^2 (v\cdot \Delta)^2
+ 2\gA \left\{v\cdot \Delta, S\cdot D\right\}\right) B
\nonumber \\
&&\ \ + \ \O\left(\frac{1}{m^2}\right).\label{oneoverm}
\eea

While Eq.(\ref{oneoverm}) is the first ``$1/m$" correction, it is not the
entire
story to the order considered. One can see that it is also the next order in
the chiral counting in derivatives and it is expected in any case,
independently of the inverse baryon mass corrections.
Thus in a practical sense, the coefficients that
appear in each term are parameters rather than fixed by chiral symmetry in
HFF. We also note that in the derivation given above, neither
$O(1/m)$ corrections to the quartic fermion term nor
sixth-order fermion terms are generated by the $(1/m)$ expansion. This
of course does not mean that such terms cannot contribute on a general ground.
We have however checked that no
such terms arise from exchanges of
single heavy mesons that are formally
``integrated out" which means that our calculations are not affected by
such terms, at least to the order we are concerned with.
\section{Counting Rules}
\indent\indent
In this section, we rederive and generalize somewhat Weinberg's counting
rule\cite{wein90} using HFF. Although we do not consider explicitly
the vector-meson degrees of freedom, we include them here in addition
to pions and nucleons. Much of what we obtain turn out to be valid in
the presence of vector mesons. Now in dealing with them, their
masses -- which are comparable to the chiral scale $\Lambda_{\chi}$
-- will be regarded as
heavy compared to the momentum probed $Q$ -- say, scale of external
three momenta or $m_\pi$.

In establishing the counting rule, we make the following key assumptions:
Every intermediate meson (whether heavy or light) carries a four-momentum
of order of $Q$.
In addition we assume that
for any loop, the effective cut-off in the loop integration
is of order of $Q$. We will be more precise as to what this means physically
when we discuss specific processes, for this clarifies the limitation of
the chiral expansion scheme.

An arbitrary Feynman graph can be characterized by the number $E_N (E_H)$
of external -- both incoming and outgoing -- nucleon (vector-meson)
lines, the number $L$ of loops, the number
$I_N (I_\pi,\ I_H)$ of internal nucleon (pion, vector-meson) lines.
Each vertex can in turn be  characterized by
the number $d_i$ of derivatives and/or of $m_\pi$ factors and
the number $n_i$ $(h_i)$ of nucleon (vector-meson) lines attached
to the vertex. Now
for a nucleon intermediate state of  momentum $p^\mu=mv^\mu + k^\mu$ where
$k^\mu = {\cal O}(Q)$, we acquire a factor $Q^{-1}$ since
\be
S_F(mv+k) = \frac{1}{v\cdot k} = {\cal O}(Q^{-1}).
\ee
An internal pion line contributes a factor $Q^{-2}$ since
\be
\Delta(q^2;m_\pi^2)=\frac{1}{q^2 - m_\pi^2} = {\cal O}(Q^{-2})
\nonumber
\ee
while a vector-meson intermediate state contributes
$Q^0$ $(\sim O(1))$ as one can see from its propagator
\be
\Delta_F (q^2;m_V^2) = \frac{1}{q^2 - m_V^2} \simeq \frac{1}{-m_V^2}
= {\cal O}(Q^0)
\ee
where $m_V$ represents a generic mass of vector mesons.
Finally a loop contributes a factor $Q^4$ because
its effective cut-off is assumed to be of order of $Q$.
We thus arrive at the counting rule that an arbitrary graph is characterized
by the factor $Q^\nu$ with
\be
\nu = - I_N - 2 I_\pi + 4 L + \sum_i d_i
\label{piN0}\ee
where the sum over $i$  runs over all vertices of the graph.
Using the identities,
$I_\pi + I_H + I_N = L + V -1$,
$I_H = \frac12 \sum_i h_i - \frac{E_H}{2}$ and
$I_N = \frac12 \sum_i n_i - \frac{E_N}{2}$,
we can rewrite the counting rule
\be
\nu = 2 - \frac{E_N + 2 E_H}{2} + 2 L + \sum_i \nu_i,\ \ \ \ \
\nu_i \equiv d_i + \frac{n_i+ 2 h_i}{2} - 2 .
\ee
We recover the counting rule derived by Weinberg \cite{wein90} if we set
$E_H=h_i=0$.

The situation is different depending upon whether or not
there is external gauge field ({\it i.e.}, electroweak field) present
in the process. In its absence (as in nuclear forces),
$\nu_i$ is non-negative
\be
d_i + \frac{n_i + 2 h_i}{2} - 2 \geq 0.
\ee
This is guaranteed by chiral symmetry \cite{wein90}. This means that the
leading order effect comes from graphs with vertices satisfying
\be
d_i + \frac{n_i + 2 h_i}{2} - 2 = 0\,.
\ee
Examples of vertices of this kind are:
$\pi^k NN$ with $k\geq 1\, (d_i=1,\ n_i=2,\ h_i=0)$,
$h N N\, (d_i=0,\ n_i=2,\ h_i=1)$, $(\Nbar \Gamma N)^2\,
(d_i=0,\ n_i=4,\ h_i=0)$,
$h\pi^k$ with $k\geq 1 \,(d_i=1,\ n_i=0,\ h_i=1)$, etc where $h$ denotes
vector-meson fields.

In $NN$ scattering or in nuclear forces, $E_N=4$ and $E_H=0$, and so we have
$\nu \geq 0$.
The leading order contribution corresponds to $\nu=0$,
coming from three classes of diagrams; one-pion-exchange,
one-vector-meson-exchange and four-fermion contact graphs.
In $\pi N$ scattering, $E_N=2$ and $E_H=0$, we have
$\nu \geq 1 $ and the  leading order comes from
nucleon Born graphs, seagull graphs and one-vector-meson-exchange
graphs.\footnote{We note here that scalar glueball fields $\chi$ play only
a minor role in $\pi N$ scattering because the $\chi \pi\pi$
vertex ($d_i=2,\ n_i=0,\ h_i=1$) acquires an additional $Q$ power.}

In the presence of external fields, the condition becomes \cite{mr91}
\be
\left( d_i + \frac{n_i + 2 h_i}{2}-2\right) \geq -1 \,.\label{exchcond}
\ee
The difference from the previous case comes from the fact that a derivative
is replaced by a gauge field. The equality holds only when
$h_i=0,\,n_i=2$ or $h_i=0,\,n_i=0$. We will
later show that this is related to the ``chiral filter" phenomenon.
The condition (\ref{exchcond}) plays an important role in determining
exchange currents.
Apart from the usual nucleon Born terms which are in the class of
``reducible" graphs and hence do not enter into our consideration,
we have two graphs that contribute in the leading order to the exchange
current: the ``seagull" graphs and ``pion-pole" graphs \footnote{These are
standard jargons in the literature. See \cite{review,chemrho}.},
both of which involve
a vertex with $\nu_i=-1$. On the other hand, a vector-meson-exchange graph
involves a $\nu_i= +1$ vertex. This is because $d_i=1,\ h_i = 2$
at the $J_\mu hh$ vertex. Therefore vector-exchange graphs are suppressed
by power of $Q^2$. This counting rule is the basis for establishing
the chiral filtering even when vector mesons are present (see Appendix I).
Thus the results
we obtain without explicit vector mesons are valid more generally.
\section{Renormalization in Heavy-Fermion Formalism}
\indent\indent
In this section, we discuss renormalization in heavy fermion formalism.
Most of the renormalized quantities that we will write down here have been
obtained by others in standard ChPT \cite{gss}. We rederive them for
completeness and as a check of our renormalization procedure.

For reasons stated above, fermion loops are suppressed in HFF.
Our basic premise is that antiparticle solutions should be
irrelevant to physical processes in large-mass and low-energy situations.
Their effects can however be systematically taken into account in
``$1/m$" expansion.

We shall denote ``bare" quantities by $\mnot,\ \Mnot,\ \Fnot,\ \gAnot$
and the corresponding ``renormalized (to their physical values)"
quantities by $m=\mnot + \delta m,\ M,\ F,\ \gA$,
respectively, for nucleon mass, pion mass, pion decay constant
($\simeq 93$ MeV) and axial coupling constant ($\simeq 1.25$).
\subsection{Dimensional regularization}
\indent\indent
We adopt the dimensional regularization scheme to handle ultraviolet
singularities in our loop calculations. It has the advantage of avoiding
power divergences like $\delta (0)\sim \Lambda^4_{cut}$ where
$\Lambda_{cut}$ is the cut-off mass. In $d= 4 - 2 \epsilon$ dimensions,
all the infinities are absorbed in $\frac{1}{\epsilon}$.
When heavy fields are involved, somewhat different
parametrization schemes and integral formulas are needed. The relevant ones
for our calculation are
\bea
\frac{1}{A^n B} &=&
2^n \frac{\Gamma(n+1)}{\Gamma(n)} \int^\infty_0 d\lambda
\frac{\lambda^{n-1}}{(2 \lambda A + B)^{n+1}}\,,
\\
\frac{1}{A^n B C} &=&
2^n \frac{\Gamma(n+2)}{\Gamma(n)} \int^1_0 d z
\int^\infty_0 d\lambda
\frac{\lambda^{n-1}}{\left[ 2 \lambda A + z B + (1-z) C\right]^{n+2}}
\eea
and
\be
\Gamma(n) \int^\infty_0 d \lambda \lambda^k (\lambda^2 + \Box^2)^{-n}
= \frac12 \Box^{1+k -2n} \Gamma\left(n - \frac{k+1}{2}\right)
\Gamma\left(\frac{k+1}{2}\right).
\ee
This integral is singular when
\be
n - \frac{k + 1}{2} = 0,\, -1,\, -2,\cdots
\nonumber\ee
so $\epsilon$ must be kept finite until the integration in
$\lambda$ is performed. Some relevant integral identities needed in this
paper are given in Appendix C.

As is customary in the dimensional regularization, we introduce an arbitrary
mass scale $\mu$. After renormalization, the results should, of course,
be independent of the scale $\mu$.
Here some comments are in order regarding the one-loop renormalization scheme.
First, all the divergences of our theory can be classified in
two classes by their degrees of divergence: quadratic and logarithmic
divergences. The quadratic
divergences are removed by counter terms that are of {\it exactly the same
form as the lowest chiral-order Lagrangian}, ${\cal L}_0$ as given by
(\ref{chiralag2}).  Thus these quadratic
singularities can be absorbed into the renormalization process of
the basic quantities, namely, $\gA,\ F,\ M,\ m$.
To take care of the logarithmic singularities, we
include counter terms that are higher chiral order by $Q^2$ than
${\cal L}_0$. As it is an arduous task \cite{gss} to write down
all possible counter terms, we shall write down only the counter terms
needed for the calculation.
All the quadratic divergences can be written
in terms of $\Delta(M^2)$ and all the logarithmic divergences
in terms of $\LL$ defined by
\bea
\Delta(M^2)&=& \frac{\mu^{4-d}}{i}\int\, \frac{d^d l}{(2\pi)^d}
\frac{1}{l^2-M^2}= - \frac{M^2}{16\pi^2} \Gamma(-1+\epsilon)
\left(\frac{M^2}{4\pi\mu^2}\right)^{-\epsilon}\nonumber\\
&=&\frac{M^2}{16\pi^2} \left( \frac{1}{\epsilon} + 1 + \Gamma'(1)
- \ln \frac{M^2}{4 \pi \mu^2}\right) + {\cal O}(\epsilon),
\label{DM2}\\
\LL &=& \frac{1}{16\pi^2}
\Gamma(\epsilon) \left(\frac{M^2}{4\pi\mu^2}\right)^{-\epsilon}
= \frac{1}{16\pi^2}
\left(\frac{1}{\epsilon}+\Gamma'(1)-\ln \frac{M^2}{4\pi\mu^2}\right)
+ {\cal O}(\epsilon)\label{LL}
\eea
where $\epsilon= (4-d)/2$ and $\Gamma'(1) \simeq -0.577215.$
Note that both $\Delta$ and $\LL$  are scale-dependent and singular.
But the coefficients of the counter terms (written down below)
are also scale-dependent and
singular. To remove the scale dependence and singularity, one must
adjust the coefficients of the counter terms. This procedure however is
not unique since finite parts of the coefficients of the counter terms are
totally arbitrary. There are many ways to eliminate this non-uniqueness
leading to a variety of ``subtraction schemes." In this paper, we use the
scheme whereby renormalization is made at the on-shell point for the
nucleon and at zero four-momentum for the pion and the current.
Thus the quantities $\gA$, $F$ and the coefficients of the counter terms
are defined at zero momentum of the axial current and the pion.

To make our discussion on renormalization streamlined, we list below {\it all}
the counter terms needed in our work (for the meson sector, see \cite{gl})
to which we will refer back as required;
\bea
{\cal L}_\ct&=& \Bbar\left[-\delta m +(Z_N-1)iv\cdot D +
i \frac{c_1}{F^2}(v\cdot D)^3 \right]B \nonumber \\
&+& i \gA \frac{c_2}{F^2} \Bbar \, S^\mu v^\nu v^\alpha
\left(\Delta_\mu D_\nu D_\alpha
- \stackrel{\leftarrow}{D_\nu} \Delta_\mu D_\alpha
+ \stackrel{\leftarrow}{D_\nu} \stackrel{\leftarrow}{D_\alpha}
\Delta_\mu \right) B
\nonumber \\
&+& i \frac{c_3}{F^2} \Bbar v^\alpha \left[ D^\beta,
\left[D_\alpha, D_\beta\right]\right] B
\nonumber \\
&+& i \frac{c_4}{F^2} \Bbar \left[S^\alpha,S^\beta\right]
\left\{ v\cdot D, \left[D_\alpha, D_\beta\right]\right\} B
\nonumber \\
&+& i \frac{c_5}{F^2} \Bbar \left[ v\cdot \Delta,\left[
v\cdot D, v\cdot \Delta\right]\right]B
\nonumber \\
&+& i \frac{c_6}{F^2} \Bbar \left[ S\cdot \Delta,\left[
v\cdot D, S\cdot \Delta\right]\right]B
\nonumber \\
&+&\left\{- \frac{\gA}{8 F^4} d_4^{(1)} \, \Bbar \tau_a D_\mu B
\cdot \Bbar \left[ v\cdot \Delta,\tau_a \right] S^\mu B
\right.\nonumber \\
&& -\, \frac{\gA}{4F^4} d_4^{(2)} \left(
\Bbar \left[S^\mu, S^\nu\right] D_\mu B \cdot \Bbar v\cdot \Delta\,
S_\nu B +
\Bbar \left[S^\mu, S^\nu\right] v\cdot\Delta\,D_\mu B \cdot \Bbar
S_\nu B\right)\nonumber \\
&&\left. +\, \  \mbox{h.c.} \right\}
\label{allct}\eea
with
\bea
Z_N  &=& 1+\frac{3(d-1) g_A^2}{4} \frac{\Delta(M^2)}{F^2},
\nonumber \\
\delta m &=& \frac{3 g_A^2 M^3}{32\pi F^2},
\nonumber \\
c_1 &=& \frac32 \gA^2 \LL + c_1^R,
\nonumber \\
c_2 &=& \frac{d-3}{3} \gA^2 \LL + c_2^R,
\nonumber \\
c_3 &=& -\frac16 \left[1+(d+1) \gA^2\right] \LL + c_3^R,
\nonumber \\
c_4 &=& - 2 \gA^2 \LL + c_4^R,
\nonumber \\
c_5 &=& \LL + c_5^R,
\nonumber \\
c_6 &=& - (d-3) \gA^4 \LL + c_6^R,
\nonumber \\
d_4^{(1)} &=& \kappa_4^{(1)} + \left[(d-1) \gA^2 -2 \right] \LL,
\nonumber \\
d_4^{(2)} &=& \kappa_4^{(2)} - 8 \gA^2 \LL
\eea
where $c_i^R \ (i=1,2,\cdots,6)$ and $\kappa_4^{(1,2)}$ are finite
renormalized constants that we will refer to as ``finite counter terms."
Since these finite counter terms and the
finite parts of loop contributions are scale-independent, our final results
also are scale-independent and regular.
The chiral counting is immediate from the counter-term
Lagrangian. We should mention for later purpose that the two-derivative
four-fermion counter terms proportional to $d_4^{(1,2)}$ cannot be gotten
from single low-mass resonance exchanges and hence do not figure in long-range
as well as intermediate-range NN potentials.

We should note here that although the above counter-term Lagrangian contains
the isospin matrix $\tau$, chiral invariance of the Lagrangian is preserved.
This can be verified by noting that
\be
\left(R^{-1} \tau_a R\right)_{ij} \left(R^{-1} \tau_a R\right)_{kl}
= \left(\tau_a\right)_{ij} \left(\tau_a\right)_{kl}
\ee
where $(i,j,k,l=1,2)$ and
$R$ is the SU(2) matrix for chiral transformation,
$B\rightarrow R B$.\footnote{Actually this equation can be simply understood
if one uses an $O(3)$ representation. In this case, $R$ is an orthogonal
real matrix and $(\tau_a)_{ij}= -i \epsilon_{aij}$, $a,i,j=1,2,3$.}
Also note that $S^\mu$ is hermitian while $\Delta_\mu$ and
$\left[S^\mu, S^\nu\right]$ are antihermitian and
$\gamma^0 D_\mu^\dagger \gamma^0 = \stackrel{\leftarrow}{D_\mu}
\equiv \stackrel{\leftarrow}{\partial_\mu} - \Gamma_\mu$
where $\Gamma_\mu$ is the antihermitian operator defined by
$D_\mu = \partial_\mu + \Gamma_\mu$.

\subsection{Pion properties to one loop}
\indent\indent
Since due to pair suppression fermion loops can be ignored,
renormalization in the pion properties
is rather simple. The wavefunction renormalization $Z_\pi$,
renormalized mass $M$ and
pion decay constant $F$ (here as well as in what follows
renormalized at $q^2=0$ with $M\neq 0$) are given by \cite{gl}\footnote{
We have not put in the counter terms that appear at $O(Q^4)$ in the
pure meson sector, so the $L_i^r$ terms of Gasser and Leutwyler \cite{gl}
are missing from our expression. Since the meson sector
proper does not concern us here, we will leave out such terms from now on.}
\bea
Z_\pi &=& 1 - \frac{2}{3} \frac{\Delta(M^2)}{F^2}, \\
M^2 &=&  \Mnot^2 \left[ 1 - \frac{\Delta(M^2)}{2F^2}
\right],
\\
F &=& \Fnot \left[ 1 + \frac{\Delta(M^2)}{F^2}\right] 
\eea
with $\Delta(M^2)$ defined in Eq.(\ref{DM2}).
\subsection{Nucleon properties to one loop}
\indent\indent
One-loop graphs for nucleon propagator are given in Fig. 3. Fig. 3$b$ vanishes
due to isospin symmetry, so
only Fig. 3$a$ survives to contribute to the nucleon self-energy $\Sigma$,
\be
\Sigma(v\cdot k) = -3 \frac{\gA^2}{F^2} S^\alpha S^\beta \int_l
\frac{l_\alpha l_\beta}{v\cdot (l+k) \, (l^2- M^2)}
\nonumber\ee
where (and in what follows)
\be \int_l \equiv \intl\,.
\ee
{}From this, we get
\be
Z_N
\,=\, 1 + \Sigma'(0)
\, = \, 1+ \frac{3(d-1)\gA^2}{4} \frac{\Delta(M^2)}{F^2}
\ee
where the prime on $\Sigma$ stands for derivative with respect to
$v\cdot k$. In our case confined to irreducible graphs, there is no
nucleon pole, so we can set $v\cdot k=0$ in the denominator of the nucleon
propagator.  For an {\it off-shell} nucleon, we have
\be
\Sigma(v\cdot k) = (d-1) \frac{3\gA^2}{4 F^2} h(v\cdot k)
\ee
with the function $h(v\cdot k)$ defined by
\be
\int_l \frac{l_\alpha l_\beta}{v\cdot (l+k) \, (l^2- M^2)}
= g_{\alpha\beta} h(v\cdot k) + v_\alpha v_\beta (\cdots)
\ee
where $(\cdots)$ stands for a function that does not concern us.
The evaluation of $h(v\cdot k)$ is described in Appendix D.
For $(v\cdot k)^2 \leq M^2$, we have
\be
\Sigma(y) = \frac{3(d-1)\gA^2}{4F^2} \Delta(M^2)\, y - \frac{3\gA^2}{2F^2}
\LL \,y^3 + \frac{3\gA^2}{4F^2} (M^2-y^2) \,{\overline{h_0}}(y),
\nonumber \ee
where $y=v\cdot k$ and $\LL$ is a singular quantity given by (\ref{LL})
and ${\overline{h_0}}$ is a finite function, the explicit form of which
is given in Appendix D.
We note that the above equation contains the $\frac{1}{\epsilon}$
divergence in
the coefficient of $(v\cdot k)^3$ as well as in the coefficient of
$(v\cdot k)$ ({\ie}, in $\Delta (M^2)$). This additional
singularity arises also in the conventional method. See the paper by
Gasser \etal \cite{gss}. The counter term needed to remove this
divergence, as given in (\ref{allct}), is
\be
\Sigma_{\ct}(y) = \delta m - (Z_N-1)\, y + \frac{c_1}{F^2} \, y^3\,.
\ee
In order to regularize the propagator subject to the
condition $\Sigma(0)=\Sigma'(0)=0$, we choose
\bea
Z_N  &=& 1+\frac{3(d-1) \gA^2}{4} \frac{\Delta(M^2)}{F^2},
\nonumber \\
\delta m &=& \frac{3 \gA^2 M^3}{32\pi F^2},
\nonumber \\
c_1 &=& \frac32 \gA^2 \LL + c_1^R.
\nonumber \eea
The result is
\be
\Sigma(y) = \delta m + \frac{c_1^R}{F^2}\,y^3 + \frac{3\gA^2}{4F^2}
\left(M^2 - y^2\right) {\overline{h_0}}(y).
\ee
Here the finite constant $c_1^R$ is in principle
to be determined from experiments. To see its
physical meaning, we should look at a process involving an off-shell nucleon.
For instance, when $v\cdot k= \pm M$, we have
\be
\Sigma(\pm M) = \pm \frac{c_1^R}{F^2} M^3 + \delta m.
\nonumber \ee

Finally, the $\frac{1}{m}$ correction is readily seen to be
\be
\delta \Sigma(k) = -\frac{1}{2m} \left[k^2-(v\cdot k)^2\right].
\ee
\subsection{Renormalization of 3-Point Vertex Functions}
\indent \indent
In this subsection, we shall calculate three-point vertex functions to
one-loop order, in particular, $J_\mu NN$ and $\pi NN$ given in Fig. 4,
where $J_\mu= A_\mu(V_\mu)$ denotes the axial vector
(vector) current.  We treat the vector current simultaneously since
some vertices involving it figure in our calculation.
Each vertex function is a sum of
contributions from tree graphs, one-loop graphs, wavefunction renormalization,
higher-order counter term insertion and $\O\left(\frac{1}{m}\right)$
corrections, if needed. The tree-graph contribution to $J_\mu NN$ is
\bea
\GANN(\mbox{tree}) &=&  \,g_A \, \tau_a \,S^\mu ,
\nonumber\\
\GVNN(\mbox{tree}) &=&  \,\frac{\tau_a}{2} \,v^\mu.
\eea
Wavefunction renormalization produces
multiplicative coefficients $Z$, {\it i.e.}, $Z_N$ for $\GANN$ and
$Z_\pi^{\frac12} Z_N$ for $\GpiNN$ etc.

Unless noted otherwise we will always set
the momentum flow of all pions and currents to be outgoing.
The current-off-shell-nucleon couplings that we will consider
are of the type
\bea
N(mv + k) &\rightarrow & N(mv+k-q) + J_a^\mu(q) ,
\nonumber \\
N(mv + k) &\rightarrow & N(mv+k-q) + \pi_a (q)
\nonumber \eea
with the relevant momenta indicated in the parentheses.

For completeness, we list in Appendix F all the contributions to the
three-point functions of each Feynman graph.
\subsubsection{Axial vertex function $\GANN$}
\indent \indent
For off-shell nucleons, we find
\be
\GANN = \gA\,\tau_a\, S^\mu \left[ 1 + \frac{\Delta(M^2)}{F^2}
+ \frac{3(d-1) \gA^2}{4F^2}\Delta(M^2) + \frac{d-3}{4} \frac{\gA^2}{F^2}
h_3(v\cdot k, v\cdot q)\right]
\ee
where $h_3(v\cdot k,v\cdot q) \equiv \frac{1}{v\cdot q}\left[
h(v\cdot k-v\cdot q)-h(v\cdot k)\right]$ is evaluated in Appendix D.
The singularity in the above equation is removed by the counter
term contribution
\be
\left[\GANN\right]_\ct = \gA\,\tau_a\, S^\mu \left(-\frac{c_2}{2 F^2} \right)
\left[ 3(v\cdot k)^2 - 3 v\cdot k\,v\cdot q + (v\cdot q)^2\right]
\ee
with
\be
c_2 = \frac{d-3}{3} \gA^2 \LL + c_2^R
\ee
where $c_2^R$ is a finite renormalized coupling constant.
Adding the counter term contribution to the loop contribution, we obtain
a renormalized axial coupling constant by $g_A= g_A(k=0,q=0)$
where $g_A(k,q)$ is defined by
\be
{\Gamma_{ANN}^{\mu,a}}^R(k,q) \equiv g_A(k,q)\,\tau_a\,S^\mu.
\ee
Physically $g_A(k,q)$ is just the axial charge form factor for the incoming
nucleon of momentum  $m v^\mu + k^\mu$ and the axial current carrying the
momentum $q^\mu$. Explicitly it is given by
\be
\frac{g_A(k,q)}{g_A} = 1 + \frac{\gA^2}{4 F^2}
{\overline h_3}(v\cdot k, v\cdot q) - \frac{c_2^R}{2F^2}
\left[ 3(v\cdot k)^2 - 3 v\cdot k\,v\cdot q + (v\cdot q)^2\right]
\ee
where ${\overline h_3}(v\cdot k,v\cdot q)$ is a finite function defined by
\be
{\overline h_3}(v\cdot k,v\cdot q) \equiv
h_3(v\cdot k,v\cdot q) + \Delta(M^2) - \frac23 \LL
\left[ 3(v\cdot k)^2 - 3 v\cdot k\,v\cdot q + (v\cdot q)^2\right]
\ee
and
\be
\gA = \gAnot\, \left[ 1 + \frac{\Delta(M^2)}{F^2}
\left( 1+ \frac{d}{2} \gA^2\right)\right] \simeq 1.25
\ee
where we have equated the renormalized $g_A$ to the experimental value.
Finally, the $\frac{1}{m}$ correction is
\be
\delta \GANN = -\frac{1}{2m} \gA \,\tau_a \, v^\mu\, (2k-q)\cdot S.
\ee

For on-shell nucleons, $v\cdot k= v\cdot q=0$ or in Breit frame,
$v\cdot k= \frac12 v\cdot q$, so we find
\be
g_A(k,q) = g_A.
\ee
{\it Note that  we have
neither momentum dependence nor $\frac{1}{m}$ corrections
for an on-shell nucleon or in Breit frame.} This means that there will
be no one-loop correction to the $\pi NN$ vertex in the exchange currents
calculated below.

Given an experimental axial charge form factor of the off-shell
nucleon, one can fix the constant $c_2^R$.
The vertex $\GpiNN$ can be obtained by a direct calculation or by means
of Lehmann-Symanzik-Zimmermann (LSZ) formulation: Both give the same result
\be
\GpiNN^R(k,q) = -i \frac{g_A(k,q)}{F} \, \tau_a \, q\cdot S.
\ee

\subsubsection{Vector vertex function $\GVNN$}
\indent \indent
The full form of $\GVNN$ is rather involved,
\be
\GVNN = \frac{\tau_a}{2} \left( v^\mu\,F_1^V + \frac{1}{m}
\left[q\cdot S, S^\mu\right] \, F_2^V - q^\mu F_3^V\right)
\ee
where
\bea
F_1^V &=& 1 -\frac{1}{F^2} \left[f_1(q^2)- \Delta(M^2)\right]
+\frac{\gA^2}{F^2} \otimes
\nonumber \\
&& \left\{(d-1)\left[ \frac34 \Delta(M^2)
-\frac14 h_3(v\cdot k,v\cdot q) - B_2(k,q)\right]
\right. \nonumber \\
&& \ \ \ \ \ \ \left.
+  2 \left[(v\cdot q)^2 -q^2\right] \frac{\partial}{\partial q^2}
B_2(k,q) + 4 v\cdot q B_1(k,q) \right\},
\nonumber \\
F_2^V &=& 4m \frac{\gA^2}{F^2} B_0(k,q),
\nonumber \\
F_3^V &=& \frac{v\cdot q}{F^2} f_3(q^2) + \frac{\gA^2}{F^2} \left[
(d+1) -2\left( (v\cdot q)^2- q^2\right) \frac{\partial}{\partial q^2}
\right] B_1(k,q)
\eea
with the functions $f_i$,  $h_i$ and $B_i$ given explicitly in Appendix
B, D and E, respectively.
Although the above equations appear to have quadratic divergences, they
actually have only logarithmic divergences as can be seen below :
\bea
F_1^V &=& 1 - \frac94 (2 v\cdot k-v\cdot q) \frac{\gA^2}{F^2}\frac{M}{6\pi}
+ \frac{\LL}{6F^2} q^2 + \frac{d+1}{6} \frac{\gA^2}{F^2} \LL q^2
\nonumber \\
&& + \frac{d-1}{2} \LL \frac{\gA^2}{F^2} \left[ 3(v\cdot k)^2 -
3 v\cdot k \, v\cdot q +(v\cdot q)^2\right] +\cdots,
\nonumber \\
F_2^V &=& - 4 m \frac{M}{16\pi} \frac{\gA^2}{F^2} + 2 m \frac{\gA^2}{F^2}
(2 v\cdot k - v\cdot q) \LL + \cdots,
\nonumber \\
F_3^V &=& \left[ 1 + (d+1)\gA^2\right] \frac{v\cdot q}{6 F^2} \LL
+ \cdots
\eea
where the ellipsis denotes finite and $O(Q^n)|_{n>2}$ terms. Quadratic
divergences disappear because of EM gauge invariance.
We see that $\GVNN(k=q=0) = \frac12 \tau_a \, v^\mu$.
The remaining logarithmic divergences are removed by the counter term
in Eq.(\ref{allct}) of the form
\bea
\left[\GVNN\right]_\ct &=& -\frac{c_1}{2F^2} \tau_a v^\mu \left[ 3(v\cdot k)^2
- 3 v\cdot q
\,v\cdot k + (v\cdot q)^2\right]
\nonumber \\
&& + \ \frac{c_3}{2F^2} \tau_a \left( q^2\, v^\mu - v\cdot q \,q^\mu\right)
\nonumber \\
&& + \ \frac{c_4}{2F^2} \tau_a \, v\cdot(2k-q)\, \left[q\cdot S, S^\mu\right]
\label{GVct} \eea
with
\bea
c_3 &=& -\frac16 \left[1+(d+1) \gA^2\right] \LL + c_3^R,
\nonumber \\
c_4 &=& - 2 \gA^2 \LL + c_4^R.
\eea
When $v\cdot k = v\cdot q=0$, we have $F_3^V = 0$ and
\bea
F_1^V(q^2) &=& 1 + \frac{c_3^R}{F^2} q^2 - \frac{q^2}{16\pi^2F^2}\left[
\frac{1+3\gA^2}{2} K_0(q^2)- 2(1+2\gA^2) K_2(q^2)\right],
\label{F1V} \\
F_2^V(q^2) &=& - \frac{\gA^2 m}{4\pi F^2} \int_0^1 dz \sqrt{M^2 - z(1-z)q^2}
+1 \label{F2V}\eea
where we have added the $\frac{1}{m}$ correction appearing
in the second term of $F_2^V$ \footnote{Although the loop contribution
to the Pauli form factor $F_2^V$ is finite and hence requires no
infinite counter term, there is a finite counter term contributing to it
which we did not -- but should -- include in our formula. As pointed out by
Bernard {\it et al.} \cite{bkkm}, the finite counter term can be considered
as coming from the $\rho$ exchange as in vector dominance picture. This point
will be addressed more precisely in \cite{pmr2}.}.
It is easy to see in (\ref{F1V}) that the counter term constant $c_3^R$
can be related directly to the isovector charge radius of the nucleon.
We will give the precise relation later. $K_i (q^2)$ are
the finite pieces of the functions $f_i (q^2)$ defined in Appendix B,
\bea
K_0(q^2) &=& \int_0^1 dz \,{\rm ln}\left[1-z(1-z) \frac{q^2}{M^2}\right],
\nonumber \\
K_2(q^2) &=& \int_0^1 dz \,z(1-z)\,{\rm ln}\left[1-z(1-z) \frac{q^2}{M^2}
\right] .
\label{K0}\eea
\subsection{Renormalization of 4-Point Vertex Function}
\indent \indent In this section, we study the 4-point vertex functions
denoted $\GpiA$ and $\GpiV$ to one loop as given in Fig. 5, corresponding
to the process
$$ N(mv+k) \rightarrow N(mv+k-q_a-q_b) + J_a^\mu(q_a) + \pi_b(q_b)$$
where the isospin components of the current and the pion are denoted by
the subscripts $a$ and $b$, respectively.
Here $v\cdot k$ represents how much off-shell the incoming nucleon is and
$v \cdot (k-q_a-q_b)$ the same for the outgoing nucleon.
For tree graphs, we have
\bea
\GpiA(\mbox{tree}) &=& -\frac{1}{2F} \eabc \tau_c \, v^\mu.
\nonumber \\
\eea
The full formulas for non-vanishing graphs for off-shell nucleons
are given in Appendix G.
Here we limit ourselves only to the on-shell nucleon case. For axial-charge
transitions, only the six graphs Fig.5 $(a)-(f)$ survive. Figures 5 $(g),(h)$
are proportional to $S_\mu$, so suppressed for the time component and
Figures 5 $(i)-(n)$ are proportional to $v\cdot S=0$. Figures 5 $(o)-(r)$
do not contribute to the axial current. (We have included them for later
purpose, see \cite{pmr2}.)

Adding the loop contributions and tree graphs with their
wavefunction renormalization constant, we have
\bea
\GpiA &=& \eabc \tau_c \, \Gamma_{\pi A}^{\mu,-}
+ i \dab \, \Gamma_{\pi A}^{\mu,+}
\nonumber \\
\eea
with
\bea
\Gamma_{\pi A}^{\mu,-} &=& - \frac{v^\mu}{2F} \left\{1-
\frac{1}{F^2}\left[ \left(1+(d-1)\gA^2\right) {\overline{f_1}}(q^2)
- 8 \gA^2 (q\cdot S)^2 f_2(q^2) - v\cdot q_a\, h_0^A(v\cdot q_a)
\right]\right\}
\nonumber \\
&& - \frac{2 \gA^2}{F^3} \left[q\cdot S, S^\mu\right] B_0(q^2)
\nonumber\\
&&+ \frac{\gA^4}{F^3} \left( \frac{1-d}{4} \left\{q_b\cdot S, S^\mu\right\}
h_4^A(v\cdot q_a) + \frac{d-3}{4}\left[q_b\cdot S, S^\mu\right]
h_4^S(v\cdot q_a)\right),
\\
\Gamma_{\pi A}^{\mu,+} &=& - (q_a+3q_b)^\mu \frac{4\gA^2}{3F^3}\left[
\frac{1-d}{4} + 2(q\cdot S)^2 \frac{\partial}{\partial q^2}\right] B_0(q^2)
+ \frac{v^\mu}{F^3} v\cdot q_a\, h_0^S(v\cdot q_a)
\nonumber \\
&&-3\frac{\gA^4}{F^3}\left(\frac{1-d}{4}\left\{q_b \cdot S, S^\mu\right\}
h_4^S(v\cdot q_a) + \frac{d-3}{4}
\left[ q_b\cdot S, S^\mu\right] h_4^A(v\cdot q_a)\right)
\eea
where $q= q_a + q_b$, ${\overline{f_1}}(q^2)= f_1(q^2) - f_1(0)$ and
$h_i^{S,A}(y) = \frac12 \left[ h_i(y) \pm h_i(-y)\right]$.
The integrals defining the functions $f_i (q^2)$ for $i=0,1,2,3$ are listed
and evaluated in Appendix B. ($h_i$'s are defined in Appendix D
and $B_0$ in Appendix E.)
The log divergences contained in these vertices are removed by the counter
term contribution,
\bea
\left[\GpiA\right]_\ct &=& -\epsilon_{abc}\, \frac{1}{F} \, \left[
\Gamma_{VNN}^{\mu,c}\right]_\ct
\nonumber \\
&& +\ \epsilon_{abc} \tau_c\, \frac{c_5}{F^2} v^\mu (v\cdot q_a)^2
\nonumber \\
&& + \ \frac{c_6}{2F^2} \frac{v\cdot q_a}{F} \left(
i\delta_{ab} \left[S^\mu, q_b\cdot S\right] -\epsilon_{abc} \tau_c\,
\left\{ S^\mu, q_b \cdot S\right\}\right)
\eea
with
\bea
c_5 &=& \LL + c_5^R, \nonumber \\
c_6 &=& - (d-3) \gA^4 \LL + c_6^R
\eea
where $c_i^R$ are renormalized finite constants listed in Eq.(\ref{allct})
and $\left[\Gamma_{VNN}^{\mu, c}\right]_\ct$ is given by (\ref{GVct}).

With {\it soft} momentum, we have $v\cdot q_a=0$ for which we obtain a
surprisingly simple expression, {\it viz},
\bea
\Gamma_{\pi A}^{\mu=0, ab} &=& -\eabc\tau_c \frac{1}{2F} F_1^V(t)
\eea
where $t\equiv q^2 = (q_a+ q_b)^2$ and $F_1^V$ is the isovector
Dirac form factor Eq.(\ref{F1V}).
The one-loop renormalization of the
$\pi A NN$ vertex corresponds to the isovector charge radius obtainable
from the form factor $F_1^V$  for which the finite counter term $c_3^R$
plays a key role.  We see that $\GpiA$ and $\GANN$ are related,
respectively, to $\GVNN$ and $\GpiV$ calculated in Appendix G.
That the $A_\mu\pi NN$ vertex for a soft pion is simply given by
$F_1^V$ has of course been understood since a long time via current algebra
and also in terms of the $\rho$-meson exchange.

Finally for the $\frac{1}{m}$ correction, one can readily
obtain the corrections to the vertices
\bea
\delta \GpiA &=& -\eabc\tau_c \frac{1}{4mF}\left(2k^\mu -q^\mu
- v^\mu(2v\cdot k-v\cdot q) + 2 \left[q\cdot S, S^\mu\right]
\right) \nonumber \\
&& -i \dab\, v^\mu\, v\cdot q_a \frac{\gA^2}{4mF},
\nonumber \\
\delta \GpiV &=& -i \dab \, v\cdot q_a \frac{\gA}{2mF} S^\mu
\label{onem}
\eea
where $k$ is the residual momentum of the incoming nucleon and
$k-q= k-q_a-q_b$ is that of the outgoing nucleon. An important point
to note from Eq.(\ref{onem})  is that
{\it for the case $v\cdot q_a=0$ and on-shell nucleons, we have no
contribution from $\frac{1}{m}$ corrections to the time component
of the axial current and the space component of the vector current}. This is
the basis for the pair suppression we will exploit in the application
to axial-charge transitions in nuclei.

The complete list of the four-point functions involving the vector and
axial-vector currents needed here and in \cite{pmr2} is given in Appendix G.
\section{Two-Body Exchange Currents}
\indent \indent
So far we have computed one-loop corrections to the graphs involving
one nucleon. They are extractable from experimental data on nucleon
properties.
In this section we incorporate the above corrections into -- and
derive -- two-body exchange currents in heavy-fermion chiral perturbation
theory. As shown previously \cite{pmr92}, the time component of the
axial-vector current (and also the space component of the vector current
\cite{pmr2}) in the long-wavelength limit is best amenable to chiral
perturbation loop calculations.\footnote{This is the crucial point in using
ChPT in nuclei that quantifies the general discussion given in Section 2.
Since this point is often misunderstood by nuclear physics
community, we would like to stress it once more although to some it may sound
obvious and repetitive. The chiral counting on which our analysis is based
is meaningful only if $\frac{Q^2}{m^2}<<1$ where $Q$ is the characteristic
momentum or energy scale involved in the process.
Therefore ChPT cannot describe
processes that involve energy or momentum scale exceeding that criterion.
This means that processes probing short-distance interactions are not
accessible by finite-order ChPT. In particular two-body currents describing
an internuclear distance $r_{12}\leq 0.6$ fm cannot be probed by the
expansion we are using. We argued before -- and will make use of the
fact  -- that there is a natural cut-off provided by short-range nuclear
interactions that go beyond the strategy of ChPT which allows a meaningful
use of the small $Q$ expansion.}
We will work out the computation to one loop
order corresponding to the next-to-leading order in the chiral counting rules
as derived in Section 4.  The process of interest is
$$N(p_1) + N(p_2) \rightarrow N(p_1') + N(p_2') + J_a^\mu(k),$$
where we have indicated the relevant kinematics with
$q_2 = p_2' - p_2$, $q_1 = p_1' - p_1$ and the energy-momentum
conservation $q_1 + q_2 + k=0$. The process is {\it soft} in the sense that
$$v\cdot k \simeq v\cdot q_i = v\cdot (p_i'-p_i) \simeq
\O\left(\frac{Q^2}{m}\right) \simeq 0$$
where $m$ is the nucleon mass or chiral scale. This kinematics markedly
simplifies the calculation. Clearly this kinematics does not
hold, say, for energetic real photons.

It is convenient to classify graphs by the current vertex involved.
The graphs that contain $J_\mu \pi NN$ play a dominant role since
$J_\mu \pi NN \sim v^\mu$ for the axial current (and $J_\mu \pi NN\sim S^\mu$
for the vector current). The graphs which contain $J_\mu NN$ (and
$J_\mu\pi\pi NN$ for the vector current) can be ignored to the relevant order
because the role of the vector and axial currents is interchanged.
In what follows, we discuss the axial
current only. The argument for the vector current goes almost in parallel
and will be detailed in \cite{pmr2}.

What we are particularly interested in is the time component of
the axial current, with axial-charge
transitions in nuclei in mind. This is where the ``pion dominance" is
particularly cleanly exhibited. The space component is also interesting both
theoretically and phenomenologically. Theoretically Gamow-Teller
transitions -- observed to be quenched -- represent the other side of the coin
relating to the chiral filter phenomenon discussed above, namely that chiral
symmetry alone or more precisely {\it soft} mechanisms associated with it
cannot make statements on this quantity \cite{mr74}. Empirically the
quenching phenomenon is
closely associated with the missing strength of giant Gamow-Teller resonances
in nuclei. Since the treatment of the space part of the axial current requires
going beyond chiral perturbation theory, we will not pursue this issue any
further in this paper.

For convenience, we define an ``axial-charge"  operator ${\cal M}$ by\footnote{
The operator ${\cal M}$ is an isovector but in what follows we will not
explicitly carry the isospin index.}
\be
{\vec A}^{\mu=0} \equiv \frac{g_A}{4 \Fp^2} \vec{\cal M}.
\ee
We decompose ${\cal M}$ into
${\cal M} = {\cal M}_{tree} + {\cal M}_{loop}$ where ${\cal M}_{tree}$
denotes the axial charge operator coming from the one-pion-exchange tree graph
({\ie}, the soft-pion term)
and ${\cal M}_{loop}$ is what comes from
loop corrections. Further we decompose ${\cal M}_{loop}$ into
\be
{\cal M}_{loop}= {\cal M}_{1\pi} + {\cal M}_{2\pi}
\ee
where ${\cal M}_{1\pi}$ denotes the loop correction to the one-pion-exchange
axial charge operator (also referred in the literature to as
``seagull graph") and ${\cal M}_{2\pi}$ the contribution from
two-pion-exchange graphs and {\it tree graphs} involving four-fermion contact
terms with counter-term insertions. We will later argue that the
latter does not contribute.  One-loop graphs involving
four-fermion contact interactions, while allowed in the relevant chiral order,
do not however contribute either.
\subsection{Results in momentum space}
\indent\indent
As stated above, the non-zero contributions to the
time component of the axial current come only
from the graphs that contain a $J_\mu \pi NN$ vertex.
The tree seagull graph supplemented with a vertex form factor -- and properly
renormalized -- leads to\footnote{We denote particle indices by $i=1,2$ without
expliciting heavy fermion fields. For instance,
$S_1$ should be understood as the spin operator sandwiched
between $\Bbar_v$ and $B_v$ of particle 1 with velocity $v$. In this section,
$q^2$ is the four-momentum squared of the pion but we are concerned with
the situation where $\frac{q_0}{|{\bf q}|} << 1$, so the static approximation
$q^2\approx -|{\bf q}|^2$ will be made in practical calculations and also
in Fourier-transforming to coordinate space later. In fact, the static
approximation is not only natural for the chiral counting but also essential
for suppression of $n$-body forces and currents for $n >2$. More on this
later.}
\be
{\vec A^\mu}_{tree} + {\vec A^\mu}_{1\pi} =
i \tauvec_1 \times \tauvec_2 \frac{g_A}{2 \Fp^2}
\, v^\mu\,q_2\cdot S_2 \, \frac{1}{\Mp^2-q_2^2} F_1^V(q_1^2)
+ (1\leftrightarrow 2)
\label{freeA}\ee
where $F_1^V$ is the Dirac isovector form factor of the nucleon.
To the order considered, there are no further corrections.
The present formalism allows us to calculate within the scheme the form factor
$F_1^V$. The tree-graph (or soft-pion) contribution corresponds to
taking $F_1^V= 1$. The difference $(F_1^V -1)$ is given by Figs. 5$(a)-(f)$.
Taking $k^\mu\rightarrow 0$, we encounter two spin-isospin operators,
\bea
{\cal T}^{(1)} &\equiv& - 2i \tauvec_1 \times \tauvec_2 \, q \cdot S_1
 + (1 \leftrightarrow 2)
\simeq i \tauvec_1 \times \tauvec_2 \, {\vec q}\cdot({\vec \sigma_1}
+ {\vec \sigma_2}),\\
{\cal T}^{(2)} &\equiv&  2 (\tauvec_1 + \tauvec_2) \, \left[q \cdot S_2,
S_1 \cdot S_2\right] + (1 \leftrightarrow 2)
\simeq i (\tauvec_1 + \tauvec_2) \, {\vec q}\cdot{\vec \sigma_1}
\times {\vec \sigma_2}
\eea
with $q^\mu\equiv q_2^\mu\simeq -q_1^\mu.$
With the help of these operators, we can rewrite (\ref{freeA}) as
\be
{\cal M}_{1\pi} = -{\cal T}^{(1)}\,\frac{1}{\Mp^2-q^2}\, \left[F_1^V(q^2)
-1\right]
\label{msea}\ee
with ${\cal M}_{tree} = -{\cal T}^{(1)}\,\frac{1}{\Mp^2-q^2}$.

As for two-pion-exchange and four-fermion contact interaction contributions,
the relevant diagrams are those given
in Fig.6$(a)-(k)$ and their symmetrized ones. Before going into any details,
one can readily see that each graph in Fig. 6 contributes a term of
at least $O(Q)$. This can be
shown both in the conventional method and in HFF by observing that their
contributions vanish if we set $M=q_i^\mu=k^\mu=0$. This assures
us that our counting rule is indeed correct. Therefore we can neglect
all the graphs proportional to $S^\mu$ since the axial-charge operator
involves $S^0\sim O(Q/m_N)$ as stated before. Figures $(f)$, $(g)$, $(h)$ and
$(j)$ belong to this class. Now Fig. $(e)$ is identically zero because of the
isospin symmetry and Fig. $(i)$ is proportional to $v\cdot S=0$.
The graph $(k)$, involving time-ordered pion propagators, are the so-called
``recoil graphs" \cite{chemrho} which as we shall argue in Section 8
will be cancelled by similar recoil terms in reducible graphs.
So we are left with only the four graphs $(a)$, $(b)$, $(c)$ and $(d)$ to
calculate.
Without any further approximation than using HFF, the full expression
of the four graphs comes out to be
\bea
{\vec A^\mu}(a) &=& -(2\tauvec_2 -i \tauvec_1\times\tauvec_2)
\, \frac{\gB}{8 \Fp^4} \,
\left(v^\mu \, q_2\cdot S\right)_1\, f_0(q_2^2),
\nonumber \\
{\vec A^\mu}(b) &=& (2 \tauvec_2 + i \tauvec_1\times\tauvec_2)
\, \frac{\gB}{8 \Fp^4} \,
\left(q_2\cdot S\, v^\mu\right)_1\,f_0(q_2^2),
\nonumber\\
{\vec A^\mu}(c) &=& (-2\tauvec_1 - 2\tauvec_2
+ i\tauvec_1\times \tauvec_2) \frac{\gB^3}{2\Fp^4}
\,\left(v^\mu\,S^\alpha\right)_1 \left( S^\beta S^\nu\right)_2
I_{\nu,\alpha\beta}(q_2),\nonumber\\
{\vec A^\mu}(d) &=& (2\tauvec_1 + 2\tauvec_2
+ i\tauvec_1\times \tauvec_2) \frac{\gB^3}{2\Fp^4}
\,\left(S^\alpha\,v^\mu\right)_1\left(S^\nu S^\beta\right)_2
I_{\nu,\alpha\beta}(q_2),
\label{Asea}\eea
with $f_0(q^2)$ given in detail in Appendix B and $I_{\mu,\alpha\beta}(q)$
defined by
\be
I_{\mu,\alpha\beta}(q)= \int_l \frac{(l+q)_\mu l_\alpha l_\beta}{v \cdot l
\, v \cdot (l+q)\,(l^2-\Mp^2)\,\left[(l+q)^2-\Mp^2\right]}. \nonumber
\ee
This integral is evaluated in Appendix E. Using the conditions $v\cdot q_i=0$
and $k^\mu\simeq 0$,
we can rewrite them in a symmetrized form
\bea
{\vec A^\mu}(a+b) &=& \frac{v^\mu}{16\pi^2} \frac{\gB}{8\Fp^4}
\left[ K_0(q^2) - 16\pi^2\LL \right] \, {\cal T}^{(1)}, \label{a1}\\
{\vec A^\mu}(c+d) &=&\! -\frac{v^\mu}{16\pi^2} \frac{\gB^3}{16\Fp^4}\left\{
\left[-(d-1) 16\pi^2\LL + 3 K_0(q^2) + 2 K_1(q^2) \right] {\cal T}^{(1)}\right.
\nonumber\\
&& \left. -8\left[ K_0(q^2)-16\pi^2\LL \right] {\cal
T}^{(2)}\right\},\label{a2}
\nonumber \\
\left[{\vec A^\mu}\right]_\ct &=&
-v^\mu\, \frac{\gB}{16 \Fp^4} \left( d_4^{(1)} {\cal{T}}^{(1)}
+ d_4^{(2)} {\cal{T}}^{(2)}\right).\label{counter}
\eea
where $q^\mu \equiv \frac12 (q_2 - q_1)^\mu$ and $\left[{\vec
A^\mu}\right]_\ct$ is
the contribution from the counter-term Lagrangian (\ref{allct}).
The $K_i (q^2)$ are finite pieces of the functions $f_i (q^2)$ defined
in Appendix B, {\it i.e.},
$$K_0(q^2) = \int_0^1 dz \,{\rm ln}\left[1-z(1-z) \frac{q^2}{\Mp^2}\right],$$
$$K_1(q^2) = \int_0^1 dz \frac{-z(1-z) q^2}{\Mp^2 - z(1-z)q^2}.$$
The expressions (\ref{a1}) and (\ref{a2}) contain singularities in $\LL$,
which are removed by the counter term contribution with
\bea
d_4^{(1)} &=& \kappa_4^{(1)} + \left[(d-1) \gB^2 -2 \right] \LL,
\nonumber \\
d_4^{(2)} &=& \kappa_4^{(2)} - 8 \gB^2 \LL
\eea
where the renormalized constants $\kappa_4$'s are finite and scale-independent.

The resulting two-body axial-charge operator including finite counter-term
contributions is
\bea
{\cal M}_{2\pi} &=& \frac{1}{16\pi^2 \Fp^2} \left\{
\left[-\frac{3\gB^2-2}{4}K_0(q^2)
- \frac12 \gB^2 K_1(q^2)\right] {\cal T}^{(1)}
+2 \gB^2 K_0(q^2) {\cal T}^{(2)}\right\}
\nonumber \\
&& - \ \frac{1}{4 \Fp^2} \left(\kappa_4^{(1)} {\cal T}^{(1)} + \kappa_4^{(2)}
{\cal T}^{(2)}\right).
\label{mab}\eea
The two-body axial-charge operator due to loop correction is then the sum of
(\ref{msea})
and (\ref{mab})
\bea
{\cal M}_{loop} = {\cal M}_{1\pi} + {\cal M}_{2\pi}.\label{totalM}
\eea
As it stands, the constants $\kappa_4$'s are the only
unknowns in the theory as they
cannot be determined from nucleon-nucleon interactions as mentioned before.
They could in principle be extracted from two-nucleon processes like
$N+N\rightarrow N+N+\pi$ but they appear as higher-order corrections and
it is inconceivable to obtain an information on these presumably small
constants from such
processes. However as argued above, we expect the constants $\kappa_4^{(i)}$
to be numerically small and what is more significant,
when we go to coordinate space as we shall do below
to apply the operator to finite nuclei, they become $\delta$ functions and
will be completely
suppressed as we discussed in Section 2. In momentum space,  such
constant terms have also to be removed as done for the celebrated
Lorentz-Lorenz effect (or more generally for the Landau-Migdal $g_0^\prime$)
in pion-nuclear scattering \cite{GEB}.
It should be stressed that {\it once the constant counter terms are removed,
no unknown parameters enter at next to the leading
order in the chiral expansion in nuclei.} It is also noteworthy that to
the order
considered, the loop contributions are renormalization-scale independent.

\subsection{Going to coordinate space}
\indent\indent
Applications in nuclear transitions are made more readily
in configuration space.
Furthermore considerations based on ranges of nucleon-nucleon interactions
which seem necessary for rendering chiral symmetry meaningful in nuclei
are more transparent in this space. Therefore we wish to Fourier-transform
the operators (\ref{a1}) and (\ref{a2}) into a form suitable for calculations
with realistic nuclear wave functions. In doing this, we will treat the
pion propagator in static approximation, namely, $q^2\approx -|{\bf q}|^2$.
In Appendix B, we show how the
highly oscillating integrals involved in the calculation can be converted into
integrals of smooth functions by performing the Fourier transform
{\it before} doing the parametric integration.
Since the spin-isospin operators ${\cal T}^{(1)}$ and ${\cal T}^{(2)}$
contain $\vec{q}$ -- which is a derivative operator in configuration space,
it is convenient to define
\bea
\T^{(1)}&=&\tauvec_1\times \tauvec_2\,\, \hat{r}\cdot(\bfsigma_1 + \bfsigma_2),
\nonumber \\
\T^{(2)}&=&(\tauvec_1+\tauvec_2)\,\, \hat{r}\cdot(\bfsigma_1\times\bfsigma_2).
\eea
Writing Eqs.(\ref{msea}) and (\ref{mab}) in coordinate space which we will
denote by $\tilde{\cal M}$ to distinguish from the momentum-space expression,
we obtain --
modulo $\delta$ function terms mentioned above -- the principal result
of this paper:
\bea
\tilde{\cal M}_{tree}(r) &=& \T^{(1)} \frac{d}{dr}\left[
- \frac{1}{4 \pi r} \e^{- \Mp r}\right],
\label{msoft} \\
\tilde{{\cal M}}_{1\pi}(r) &=& c_3^R \frac{\Mp^2}{F^2}\,\tilde{{\cal M}}_{tree}
\nonumber \\\
&+&\frac{\T^{(1)}}{16\pi^2 \Fp^2}
\frac{d}{dr} \left\{
 - \frac{1+3 \gB^2}{2}\left[K_0(r)- {\tilde K_0}(r)\right]
 + (2+4\gB^2) \left[K_2(r) - {\tilde K_2}(r) \right]\right\},\nonumber\\
\label{onepiloop}\\
\tilde{{\cal M}}_{2\pi}(r) &=& \frac{1}{16\pi^2 \Fp^2} \frac{d}{dr}\left\{
-\left[\frac{3\gB^2-2}{4} K_0(r) +\frac12 \gB^2 K_1(r)\right] \T^{(1)}
+ 2\gB^2 K_0(r)\T^{(2)}\right\},
\label{twopiloop} \\
\tilde{{\cal M}}_{loop}(r) &=& \tilde{{\cal M}}_{1\pi}(r)
+\tilde{{\cal M}}_{2\pi}(r).
\label{mloop}
\eea
As defined, ${\cal M}_{n\pi}$ are $n\pi$ exchange corrections to the
soft-pion (tree) term. ${\cal M}_{loop}$ is therefore the total loop
correction we wish to calculate.
The explicit forms of the functions $K_i(r)$ and ${\tilde K}_i(r)$
are given in Appendix B.

As noted above, the constant $c_3^R$ can be extracted from the isovector
Dirac form factor of the nucleon, {\it i.e.},
\be
c_3^R \frac{\Mp^2}{\Fp^2} = \frac{\Mp^2}{6} \langle r^2\rangle_1^V
\simeq 0.04784.\label{c3r}
\ee
It is interesting to separate what we might call ``long wavelength
contribution" from  $\tilde{\cal M}_{1\pi}$,
$$ \tilde{\cal M}_{1\pi} (r) = \delta_{soft} \, \tilde{\cal M}_{tree}(r)
\, + \, \mbox{(short range part)} $$
where
\be
\delta_{soft} = c_3^R \frac{\Mp^2}{\Fp^2}
+ \frac{\Mp^2}{16\pi^2 \Fp^2}\left[ \frac{1+3\gB^2}{2}
\left(2-\frac{\pi}{\sqrt{3}}\right) - (1+2\gB^2) \left(
\frac{17}{9} - \frac{\pi}{\sqrt{3}} \right)\right]
\simeq 0.051
\ee
and compare this one-loop prediction for $\delta_{soft}$ \footnote{It
is worth noting that this contribution is generic in the sense that
it is more or less model-independent: It is of the same form and magnitude
whether it is given by chiral one-loop graphs or by the vector dominance
(see Appendix I) or
by the phenomenological dipole form factor.  This may have to do with the
fact that it is controlled entirely by chiral symmetry. It is curious though
that this longest wavelength effect is an {\it enhancement}
rather than quenching usually associated with form factors.}
to what one would expect from the phenomenological dipole form factor
\be
F_1^V(q^2) = \left(\frac{\Lambda^2}{\Lambda^2 - q^2}\right)^2
\ee
with $\Lambda= 840$ MeV. This form factor leads to the following one-pion
exchange contribution to ${\cal M}_{loop}$, corresponding to (\ref{onepiloop}):
\bea
\tilde{{\cal M}}_{1\pi}^{dipole} &=& \T^{(1)} \left\{ \frac{\Mp^2}{4\pi x_\pi}
Y_1(x_\pi)\left[
\left(\frac{\Lambda^2}{\Lambda^2-\Mp^2}\right)^2 -1\right]\right. \nonumber \\
&&\left.\ \ - \, \frac{1}{4\pi} \left[ \frac12
\left(\frac{\Lambda^2}{\Lambda^2-\Mp^2}\right) \Lambda^2 +
\left(\frac{\Lambda^2}{\Lambda^2-\Mp^2}\right)^2 \left(\frac{1}{r^2}
+ \frac{\Lambda}{r}\right)\right] \e^{-\Lambda r}\right\}.\label{vdm}
\eea
Identifying the first term of (\ref{vdm}) with the first term of
(\ref{onepiloop}), we see that $\delta_{soft}$ corresponds to
$$\left(\frac{\Lambda^2}{\Lambda^2-\Mp^2}\right)^2 -1\simeq 0.0571.$$
It is remarkable -- and pleasing -- that the one-loop calculation of
$\delta_{soft}$ is so close to the empirical value. Furthermore the remaining
term in (\ref{onepiloop}) involving the functions $K^\prime_i$ corresponds --
and when applied to the process of interest, is numerically close --
to the second (short-ranged) term in (\ref{vdm}).
\section{Numerical Results}
\indent\indent
In order to get a qualitative idea of the size involved, let us first
look at the magnitude of the relevant terms given in momentum
space. For this purpose we set $q^2\approx -|{\bf q}|^2\sim -Q^2$, where
$Q$ is taken to be a characteristic small momentum scale probed in the
process which we take to be of order of $m_\pi$ at most.
For convenience we shall factor out the tree contribution
from the expression (\ref{totalM}) and write it as
\be
{\cal M}={\cal M}_{tree} (1+\delta_{\mbox{\tiny M}}+O(Q^3))
\label{MtinyM}\ee
where $\delta_{\mbox{\tiny M}}$ is the chiral correction of $O(Q^2)$
that we have computed (relative to the tree contribution).
We obtain
$$\delta_{\mbox{\tiny M}} = \delta_{1\pi} + \delta_{2\pi}$$
where, setting $\langle {\cal T}^{(1)}\rangle =
\langle {\cal T}^{(2)}\rangle$ in nuclear medium
\footnote{ One can show in fermi-gas model, Wigner's $SU(4)$
supermultiplet model or
even jj-coupling shell model of nucleus with one particle outside of closed
core, $\langle {\cal T}^{(1)}\rangle =\langle {\cal T}^{(2)}\rangle$.
This relation will be assumed in all numerical calculations that follow.
We would like to thank Kuniharu Kubodera for his help on this
relation.} and dropping the $\kappa_4$'s \footnote{Dropping the constant
terms in momentum space is not fully justified unless all other terms of
the same nature are removed as well. This problem is avoided in
coordinate space. We give only the absolute values for the $\delta_{n\pi}$
for the same reason. See below for more on this matter.}
\bea
\delta_{1\pi}&\approx & \frac{Q^2}{4 \Fp^2}\left[-\frac23 \Fp^2 \langle r^2
\rangle_1^V +\frac{1+3\gB^2}{8\pi^2} K_0 (Q^2)
-\frac{1+2\gB^2}{2\pi^2} K_2 (Q^2)\right],\nonumber\\
\delta_{2\pi} &\approx & -\frac{Q^2+\Mp^2}{4 \Fp^2}\left[\frac{5\gB^2+2}{16
\pi^2} K_0 (Q^2) -\frac{\gB^2}{8\pi^2} K_1 (Q^2)\right].
\label{d1d2}\eea
For $Q\sim m_\pi\approx 140$ MeV, $\gB=1.25$ and $\Fp\approx 93$
MeV, we get
\bea
|\delta_{1\pi}| \sim 0.045,
\ \ \ \ \ |\delta_{2\pi}|\sim 7.5
\times 10^{-3}.
\eea
This is consistent with the notion that at the relevant scale $Q$, the
chiral correction remains {\it small}.

We now turn to a more realistic estimate of the chiral correction appropriate
to the actual  situation in finite nuclei.
Calculating nuclear transition matrix elements in momentum space
is cumbersome and delicate. There are several reasons for this. The most
serious problem is the implementation of the short-range correlation. In
the well-studied case as in the $\pi$-nuclear scattering, we know how to
proceed, obtaining the celebrated Lorentz-Lorenz effect.
Roughly the argument goes as follows \cite{GEB}. Consider a term of the form
$\vec{q}^2/(\vec{q}^2+\Mp^2)$ that figures in the p-wave pion-nuclear
scattering amplitude, or more specifically in the interaction between
the particle-hole
states excited by the pion. Rewrite this as $1-\Mp^2/(\vec{q}^2+\Mp^2)$.
Removing the constant 1 corresponds to suppressing a $\delta$ function
in coordinate space and leads to the Lorentz-Lorenz factor. Note that this
procedure of accounting for short-range correlations can even change the sign.
Unfortunately our case does not lend itself to a simple treatment of this
kind because of the nonanalytic terms coming from the loop contributions:
there is no economical way of
``removing $\delta$ functions" from them. This task is much simpler
and more straightforward in coordinate space.

Let us therefore turn to the coordinate space operators (\ref{msoft})
and (\ref{mloop}).
In Fig. 7, we plot  ${\tilde {\cal M}}_{tree}$ (\ref{msoft}) and
${\tilde {\cal M}}_{loop}$ (\ref{mloop}) as function of the internuclear
distance $r=|\vec{r}_1-\vec{r}_2|$ setting $\T^{(1)}=\T^{(2)}=1$.
Some of the important features discussed in the preceding sections
can be seen in this plot. While negligible at large distance, say,
$r > 1$ fm, the loop corrections get progressively
significant at shorter distances and
at $r \sim 0.4$ fm, they are comparable to the soft-pion result.
There is nothing surprising or disturbing about this feature at short
distances. At shorter distances which are probed by the momentum scale
approaching the chiral scale, there is no reason to ignore the
degrees of freedom {\it integrated out} from the theory. Low-order
calculations with higher chiral-order degrees of freedom eliminated
cannot possibly describe the short-distance physics properly. This
may be construed as a sign that ChPT is not predictive in nuclei.
We claim that this is not so. The point is that
as long as the scale $Q$ probed by experiments is much less than the
chiral scale, truncating higher chiral-order and shorter wavelength
degrees of freedom as done in ChPT
can be meaningful provided short-range nuclear correlations
are implemented in the way discussed above.

Calculations of the nuclear matrix elements with sophisticated wave functions
in finite (light and heavy) nuclei -- and comparison with experimental data --
will be made and reported in a
separate paper. Here for our purpose of getting a semi-quantitative idea,
the fermi-gas model as used by Delorme \cite{delorme} will  suffice.
One could incorporate accurate correlation
functions -- and this will be done for specific transitions in finite nuclei.
Here we will not do so. We shall instead take
the simplest correlation function, namely ${\hat g}(r,d)= \theta(r-d)$
with the cut-off distance $d\approx 0.7$ fm as used by Towner \cite{towner}.
Since this is a rather crude approach,  we will consider the range of $d$
values between $0.5$ and $0.7$ fm.

Specifically we are interested in the ratio of the matrix elements
$\langle {\cal M}_{loop}\rangle/\langle {\cal M}_{tree}\rangle$
which in fermi-gas model takes the form (see Appendix H)
\be
R(d,\rho) \equiv \frac{\langle {\cal M}_{loop}\rangle}{
\langle {\cal M}_{tree}\rangle}
= \left. \frac{\int_d^\infty dr\,r\left[j_1(\pF r)\right]^2
{\tilde {\cal M}}_{loop}(r)}{
\int_d^\infty dr\,r\left[j_1(\pF r)\right]^2
{\tilde {\cal M}}_{tree}(r)}
\right.
\ee
where
$\pF$ and
$\rho=\frac{2}{3\pi^2} \pF^3$ are, respectively,
the fermi-momentum and density of the system,
$j_1(x)= \frac{\sin x}{x^2} - \frac{\cos x}{x}$
and $\tilde{\cal M}_{loop}(r) \equiv \tilde{\cal M}_{1\pi}(r) +
\tilde{\cal M}_{2\pi}(r)$.
Note that $w(\pF,r) \equiv 4 \pi r\left[j_1(\pF r)\right]^2 / \pF^2$ can be
viewed as
a weighting function.  Since this calculation is straightforward, we
shall not go into details here. For completeness, however,
we sketch the calculation in Appendix H.

In Fig. 8 are plotted the functions $w(\pF,r) \tilde{\cal M}_{tree}(r)$ and
$w(\pF,r) \tilde{\cal M}_{loop}(r)$ with $\T^{(1)}=\T^{(2)} =1$
for $\pF\simeq 1.36$ fm$^{-1}$ corresponding to
nuclear matter density. The ratio $R(d,\rho)$ is plotted  in Figure 9
for $d= 0.5, \ 0.7$ fm. For $d=0.7$ fm which was used by
Towner\cite{towner}, the loop
correction is at most of the order of 10\% of the soft-pion term
at nuclear matter density. There are two important points to note in the
result. The first is that {\it separately} the loop corrections to the one-pion
term ({\ie}, ${\cal M}_{1\pi}$) and the two-pion term ({\ie},
${\cal M}_{2\pi}$) can be substantial but the {\it sum} is small.
The second point
is that the resulting loop contribution has a remarkably weak
density dependence. The first is a consequence of chiral symmetry reminiscent
of the tree-order cancellation in linear $\sigma$ model
of the nucleon pair term and the
$\sigma$-exchange term in the S-wave $\pi N$ scattering amplitude.
The second observation has a significant ramification on the mass dependence of
axial-charge transitions in heavy nuclei to which we will return shortly.
\section{Other Contributions}
\indent\indent
Here we briefly discuss what other graphs could potentially contribute
and the reason why they are suppressed in our calculation. Consider the
two-body graphs given by Figure 6$(k)$ where the pion propagators
are {\it time-ordered}.
They belong to what one calls ``recoil graphs" in the literature
\cite{chemrho}. To $O(Q^2)$ relative to the soft-pion term, these graphs --
and more generally {\it all} recoil graphs including one-pion exchange --
do not contribute. The reason is identical to the suppression of three-body
forces as discussed by Weinberg\cite{wein92}: the graphs in Fig. 10 are
exactly cancelled by the recoil corrections to the iterated one-pion exchange
graphs that are included in the class of reducible graphs. Thus to the extent
that the static approximation is used in defining the one-pion
exchange potential, these graphs should not be included as corrections.
Incidentally this justifies the standard practice of ignoring
recoil graphs in calculating exchange contributions in both weak and
electromagnetic processes in nuclei.

We have ignored in our calculation three-body and higher-body contributions
such as Figure 11. The reason for ignoring these graphs is identical to that
used for proving the suppression of three-body and other multi-body forces
\cite{wein92}.
As in nuclear forces, they can contribute
at $O(Q^3)$ relative to the soft-pion term\cite{ordonez}.

An interesting question to ask is in what situations the approximations
that justify dropping the graphs considered here {\it break down} in nuclei.
It is
clear that the static approximation -- one of the essential ingredients
of the heavy-fermion formalism -- must break down when the energy transfer
involved is large. Imagine that one is exciting a $\Delta$ resonance in nuclei
by electroweak field. The energy transfer is of the order of 300 MeV, so
the static approximation for the pion propagator involving a $\pi \Delta N$
vertex cannot be valid. In such cases, one would expect that multi-body forces
and currents suppressed in this work could become important. This suggests
specifically that in electron scattering from nuclei with sufficiently large
energy transfer, $n$-body currents (for $n>2$) will become progressively
more important in heavier nuclei. Combined with the dropping mass effect
({\ie}, ``Brown-Rho" scaling mentioned below), one expects a large deviation
from the standard mean-field description used currently.

\section{Conclusions and Discussions}
\indent\indent
We have used heavy-baryon chiral perturbation formalism to calculate
the leading corrections to the soft-pion axial-charge operator in
nuclei. Exploiting short-range suppression of the counter terms and other
short-range components of the two-body operator, we have
shown that the chiral filter mechanism holds in nuclear matter
with a possible uncertainty of no more than 10\%,
thus confirming the dominance of the soft-pion exchange. In a separate
paper, we will show that the same holds in electromagnetic responses
in nuclei. {\it Since the currents (both vector and axial-vector) are
calculated
consistently with the symmetries involved, they are fully consistent with
nuclear forces that are calculated to
the same chiral order}: Ward-Takahashi identities will be formally satisfied
although in practice approximations made for calculations may disturb them.
The final consistency will of course have to be checked {\it \`{a} posteriori}
case-by-case.

Taking this result as a statement of chiral symmetry of QCD in nuclei,
what can one learn from this concerning the phenomenological models popular
in nuclear physics where one uses exchanges of all the low-lying bosons
in fitting nucleon-nucleon scattering (such as the Bonn potential)
as well as calculating the exchange currents? Suppose we denote
the axial-charge two-body operator from
one-pion exchange with form factors by $A_{1\pi}$,
one-heavy-meson exchange with form factors by $A_{H}$,
the axial current form factor by $A_{FF}$, all calculated within a
phenomenological model, then our result implies that
for the model to be consistent
with chiral symmetry, then the total {\it must} sum to
\be
A_{total} = A_{1\pi}+A_{H}+A_{FF}+\cdots\approx A_{soft} (1+ \delta),
\ \ \ \ \ |\delta| << 1\label{chiralfilter}
\ee
where $A_{soft}$ is the soft-pion term as defined in this paper and
$\delta$ is the next-to-leading term of $O(Q^2)$. Our calculation illustrates
how individually significant terms conspire to give a small $O(Q^2)$
correction which is insensitive to nuclear density.

One other outcome of our result is that while a subset of graphs can
have a substantial density dependence, the small net chiral correction
from the totality of the graphs
does not have an appreciable nuclear density dependence, at least in fermi-gas
model. We see no reason why this weak density dependence should not
persist in more realistic nuclear models. Thus assuming that
$n$-th order chiral corrections for $n\geq 3$ (relative to the leading
soft-pion term) are not anomalously large, we come to
the conclusion that meson-exchange axial-charge contributions to
nuclear matrix elements cannot be substantially enhanced in heavy nuclei
over that in light nuclei. The question arises then as to what could be
the explanation for
Warburton's recent observation that while the mesonic effect is about
50\% in light nuclei, it is required to be 100\% in heavy nuclei
such as in lead region \cite{warburton}.
One suggestion \cite{kr} was that the parameters of the basic chiral Lagrangian
have to be modified in the presence of nuclear matter consistent with
trace anomaly of QCD \cite{br91}. It predicted that hadron masses and pion
decay constants that appear in the single-particle and one-pion exchange
two-body operators are scaled by a universal factor $\Phi$
that depends on matter density.
Another suggestion \cite{towner,riska} was that exchanges
of heavy mesons $\sigma$, $\rho$, $\omega$, $a_1$ etc could become important
in heavy nuclei while relatively unimportant in light systems.
The latter mechanism
relied on nucleon-antinucleon pair terms in phenomenological Lagrangians.
Both mechanisms seemed to qualitatively account for the enhancement.

We wish to understand the possible link, if any, between the chiral Lagrangian
approach and the phenomenological approach that includes pair terms involving
heavy mesons.
Since within the chiral approach developed in this paper the pair is naturally
suppressed as required by chiral symmetry and multi-body currents
are also suppressed as discussed above, the scaling mass effect of
\cite{kr,br91} is the only plausible mechanism left within low-order chiral
expansion
for the medium enhancement noted by Warburton. Needless to say, we cannot rule
out -- though we deem highly unlikely --
the possibility that higher order chiral terms supply the needed density
effect. Incorporating the possible 10\% loop correction calculated above in the
two-body operator and the scaling factor $\Phi=m_N^*/m_N\approx 0.8$,
one gets in the scheme of Ref.\cite{kr} the enhancement
in heavy nuclei (at nuclear matter density) $\epsilon_{\mbox{\tiny MEC}}
\approx 2.1$ which is reasonable in the lead region compared with the
experimental value $2.01 \pm 0.05$
\cite{warburton}. Within the scheme, this is the entire story and a
surprisingly simple one. Of course
more detailed finite nuclei calculations will be needed to make a truly
meaningful test of the theory.

In the phenomenological approach studied by the authors in \cite{towner}
and \cite{riska}, there is no fundamental reason to suppress $N\bar{N}$
pairs, so that heavy mesons could contribute through the pair term. However
the exchange of heavy mesons, particularly that of vector mesons, is suppressed
by short-range correlations in nuclear wave functions. Furthermore
in the model of Towner\cite{towner}, a large cancellation takes place in the
sum such that the relation (\ref{chiralfilter}) seems to hold
well\cite{wt92}: Towner finds $\delta < 10\%$ over a wide range of nuclei.
It is naturally tempting
to suggest that Towner's model gives a result close to ours {\it because
it is consistent with chiral symmetry,} at least to the same order of
chiral expansion as ours, with higher-order terms implicit in Towner's model
which need not be consistent with ChPT somehow cancelling out\footnote{
The following observation may be relevant to our argument that the counter
terms $\kappa_4^{(i)}$ {\it must} be ignored. Suppose one constructs a
purely phenomenological theory based on meson-exchange picture by fitting
experiments but conform to the symmetries of the strong interactions. Towner's
model is one such example. One can convince oneself that in such a model,
it is not possible to generate counter terms of the $\kappa_4$ type in
infinite mass limit. Therefore if such terms existed, then they must be due
to degrees of freedom that are {\it not relevant} at the accuracy required.}.
There is however one aspect that needs to be clarified:
in the models of \cite{towner} and \cite{riska}, there is a pair term
associated with a scalar meson ($\sigma$) exchange. In the chiral Lagrangian
used in this paper, there is no equivalent scalar field. We have however the
scalar field $\chi$ associated with the trace anomaly of QCD which
plays a role in the Brown-Rho scaling\cite{br91}. We believe that these
two effects are roughly related in the sense discussed in Ref.\cite{brownetal}.
In this sense we would say that the pair term involving  the $\sigma$ meson
is simulating the density-dependence of the nucleon mass in the {\it
one-body} axial-charge operator. There is no mechanism in
Ref.\cite{towner,riska}, however,
for the density dependence of Ref.\cite{kr} in the two-body operator.
We suggest that this can be generated by taking three-body terms with
an $N\bar{N}$ pair coupled to a $\sigma$-exchange.

An obvious omission in our treatment of the axial current is the space
component of the current governing
Gamow-Teller transitions in nuclei (and the time part of the
electromagnetic current in \cite{pmr2}). The reason for this was already
stated at several points in the paper: this part of the
current is {\it not} dominated by
a soft-pion exchange and indeed as noted many years ago \cite{mr74}
it is rather the very short-ranged part of nuclear interactions (roughly
equivalent to the removal of the $\delta$ functions associated with
the counter terms in the spin-isospin channel) that plays an
important role, {\it e.g.} in quenching the axial-vector coupling constant
from the free-space value $\gB=1.25$ to $\gB^*\approx 1$ in nuclear matter.
(For a similar situation with the isoscalar axial-charge transition
mediated by a neutral weak current where soft-pion exchange is forbidden,
see Ref.\cite{riskaetal}.) Furthermore, three-body operators for the space
component of the axial current may not be negligible.
For instance, as one can see from Appendix A, the three-body Gamow-Teller
operator involving one nucleon with an $A_\mu \pi\pi NN$ vertex with the pions
absorbed by two other nucleons is not trivially suppressed as it is for
the axial-charge operator.
This suggests that low-order chiral perturbation theory may have little
to say about this aspect of nuclear interactions. It is intriguing that
in nuclei, both chiral and non-chiral aspects of QCD seem to coexist in the
same low-energy domain. This makes QCD in nuclei quite different from
and considerably more intricate than QCD in elementary particles studied by
particle physicists.

Finally we mention a few additional issues we have not treated in this paper
but we consider to be important topics for future studies.
\bitem
\item It would be interesting to see what two-loop (and hopefully higher-loop)
and corresponding chiral
corrections do to the chiral filter phenomenon. Two-loop calculations are
in general a horrendous task but the situation in nuclear
axial-charge transitions might be considerably simpler than in other
processes.
\item It would be important to see whether ChPT is predictive for processes
involving larger momentum transfers as well as large energy transfer.
{}From our experience
with the electrodisintegration of the deuteron at large momentum transfers
where the naive soft-pion approximation seems still to work fairly well, we
conjecture that the chiral filter mechanism holds still in some channels
even in processes probed at shorter distances or at larger momentum transfers.
But as mentioned above, large energy transfer electron scattering might
require multi-body currents in heavy nuclei.
\item In this paper, we worked with an effective Lagrangian in which {\it
all other degrees of freedom} than pions and nucleons have been integrated out.
It would be important to reformulate ChPT using a Lagrangian that contains
vector mesons incorporated \`{a} la hidden gauge symmetry (HGS)
\cite{bando}\footnote{As stressed by Georgi, the HGS is an approach most suited
to a systematic chiral counting when vector mesons are explicitly present.
See \cite{vectorlim}.}
and also nucleon resonances (such as $\Delta$). As mentioned before, we
believe that the chiral filter argument presented in this paper is not
modified in the presence of these resonances in the Lagrangian. In Appendix I,
we show that the presence of vector mesons does not modify our prediction
on the axial-charge operator. Furthermore
it is not difficult to see that the baryon
resonances -- in particular the $\Delta$ resonances -- do not contribute
to the axial-charge transitions to the order considered. However
as is known for Gamow-Teller transitions in nuclei \cite{mr74}, certain
processes in nuclei might require, even at zero momentum transfer,
an explicit role of some of these heavier particles. As recently shown by
Harada and Yamawaki \cite{yamawaki}, vector mesons introduced via HGS can
easily
be quantized, so their implementation in ChPT would pose no great difficulty.
\item A systematic higher-loop
chiral perturbation approach using the same heavy-fermion formalism
to kaon-nuclear interactions
and kaon condensation has not yet been worked out. This is an important issue
for hypernuclear physics, relativistic heavy-ion physics and stellar collapse
\cite{brt92}.
\item Finally if the parameters of effective chiral Lagrangians scale as a
function of nuclear matter density as suggested by Brown and Rho\cite{br91},
then one expects that as matter density increases, many-body currents will
become increasingly important even at small energy transfer.
This was already noticed in \cite{kr}
where the soft-pion exchange charge operator became stronger in heavier
nuclei. We already noted that this effect will show up more prominently
in nuclear electromagnetic
responses with large energy transfer.
Future accurate experiments in electron scattering off nuclei will
test this prediction.  Of course this issue has to be treated together
with many-body forces that enter into such processes.
\eitem

\subsubsection*{Acknowledgments}
\indent\indent
We are grateful for useful discussions with G.E. Brown, K. Kubodera,
U.-G. Meissner, D.O. Riska, I.S. Towner and E.K. Warburton.
One of us (DPM) wishes to acknowledge the hospitality of Service de
Physique Th\'{e}orique of CEA Saclay where part of his work was done.
The work of
TSP and DPM is supported in part by the Korean Science and Engineering
Foundation through the Center for Theoretical Physics, Seoul National
University.


\pagebreak

\newpage
\section*{Appendix A: Chiral Lagrangian Eq.(\ref{chiralag2}) Expanded}
\renewcommand{\theequation}{A.\arabic{equation}}
\setcounter{equation}{0}

For completeness we expand ${\cal L}_0$ in powers of the pion field
and in external fields. We will group ${\cal L}_0$ by the number of
external gauge field, ${\cal L}_0 = {\cal L}_0^0 + {\cal L}_0^1 +
{\cal L}_0^2$,
\bea
{\cal L}_0^0
&=& \frac12 (\dmu \pivec)^2 - \frac12 M^2 \pivec^2 - \frac{1}{6F^2}
\left[\pivec^2 \dmu\pivec\cdot \dmup \pivec - (\pivec\cdot \dmu\pivec)^2\right]
+ \frac{M^2}{4! \, F^2} \left(\pivec^2\right)^2
\nonumber \\
&&+\  \Bbar i v\cdot \partial B
\nonumber \\
&&+\ \Bbar \left\{-\frac{v^\mu}{4F^2} \tauvec\cdot\pivec\times\dmu\pivec
- \frac{\gA}{F} S^\mu \tauvec\cdot\left[\dmu\pivec
+ \frac{1}{6F^2}\left(\pivec \,\pivec\cdot\dmu\pivec- \dmu \pivec\,
\pivec^2\right)\right]\right\}B
\nonumber \\
&&-\ \frac12 C_a \left(\Bbar \Gamma_a B\right)^2 + \cdots,
\\
{\cal L}_0^1
&=& \vec{\cal V}^\mu \cdot\left[ \pivec\times\dmu\pivec -\frac{1}{3F^2}
\pivec\times\dmu\pivec\,\pivec^2\right]
- F \vec{\cal A}^\mu\cdot\left[\dmu\pivec -\frac{2}{3F^2}\left(\dmu\pivec
\,\pivec^2 - \pivec\,\pivec\cdot\dmu\pivec\right)\right]
\nonumber \\
&&+\ \frac12 \Bbar \left(v^\mu \vec{\cal V}_\mu+2 \gA S^\mu \vec{\cal A}_\mu
\right)\cdot \left[\tauvec + \frac{1}{2F^2}\left(\pivec\,\tauvec\cdot\pivec
-\tauvec\,\pivec^2\right)\right] B
\nonumber \\
&&+\ \frac12 \Bbar \left(v^\mu \vec{\cal A}_\mu+2 \gA S^\mu \vec{\cal V}_\mu
\right)\cdot \left[\frac{1}{F}\tauvec\times\pivec - \frac{1}{6F^3}
\tauvec\times\pivec\,\pivec^2\right] B + \cdots,
\\
{\cal L}_0^2
&=& \frac12 F^2 {\vec{\cal A}_\mu}^2 + F \pivec\cdot \vec{\cal V}_\mu
\times {\vec{\cal A}}^\mu + \frac12 \left[ \pivec^2 \left(
{\vec{\cal V}_\mu}^2- {\vec{\cal A}_\mu}^2 \right)
- (\pivec\cdot \vec{\cal V}_\mu)^2  + (\pivec\cdot \vec{\cal A}_\mu)^2 \right]
+ \cdots .
\eea

{}From the (A.2), we extract Noether currents,
\bea
{\vec A}^\mu &=&
- F \left[\dmu\pivec -\frac{2}{3F^2}\left(\dmu\pivec
\,\pivec^2 - \pivec\,\pivec\cdot\dmu\pivec\right)\right]
\nonumber \\
&&+\ \frac12 \Bbar \left\{2 \gA S^\mu
\left[\tauvec + \frac{1}{2F^2}\left(\pivec\,\tauvec\cdot\pivec
-\tauvec\,\pivec^2\right)\right]
+ v^\mu \left[\frac{1}{F}\tauvec\times\pivec - \frac{1}{6F^3}
\tauvec\times\pivec\,\pivec^2\right] \right\}B + \cdots,\nonumber\\
\\
{\vec V}^\mu &=& \left[ \pivec\times\dmu\pivec -\frac{1}{3F^2}
\pivec\times\dmu\pivec\,\pivec^2\right]
\nonumber \\
&&+\ \frac12 \Bbar \left\{v^\mu
\left[\tauvec + \frac{1}{2F^2}\left(\pivec\,\tauvec\cdot\pivec
-\tauvec\,\pivec^2\right)\right]
+2 \gA S^\mu
\left[\frac{1}{F}\tauvec\times\pivec - \frac{1}{6F^3}
\tauvec\times\pivec\,\pivec^2\right]\right\} B + \cdots.\nonumber\\
\eea

\section*{Appendix B: Functions $f_i(q^2)$}
\renewcommand{\theequation}{B.\arabic{equation}}
\setcounter{equation}{0}
\indent\indent

The functions  $f_i(q^2)$ $(i=0,1,2,3)$ figuring in subsections (5.4.2)
and (5.5) are defined
by
\bea
f_0(q^2) &=&
\int_l \frac{1}{(l^2-M^2)\,\left[ (l+q)^2 - M^2\right]},
\nonumber\\
\frac12 g_{\alpha\beta} f_1(q^2) + q_\alpha q_\beta f_2(q^2)
&=& \int_l \frac{(l+q)_\alpha l_\beta}{(l^2-M^2)\,
\left[(l+q)^2 - M^2\right]}
\nonumber\\
f_3(q^2) &=& 2 f_2(q^2) + \frac12 f_0(q^2)
\eea
where we have defined $f_3(q^2)$ through
\be
\int_l \frac{l_\alpha\, (2l+q)_\beta}{(l^2-M^2)\, \left[(l+q)^2-M^2\right]}
\equiv g_{\alpha\beta}f_1(q^2) + q_\alpha q_\beta f_3(q^2).
\ee
Here and in what follows, the mass $M$ could be thought of as the pion mass
$\Mp$.
One can verify that
$$\frac{\partial f_1(q^2)}{\partial q^2} = f_2(q^2). $$
After performing the parametric integration, we have
\bea
f_0(q^2) &=& \LL - \frac{1}{16\pi^2} K_0(q^2),
\nonumber \\
f_1(q^2) &=& \Delta(M^2) -\frac{\LL}{6} q^2 - \frac{q^2}{16\pi^2}
\left[ 2 K_2(q^2) - \frac12 K_0(q^2) \right],
\nonumber \\
f_2(q^2) &=& -\frac{\LL}{6} + \frac{1}{16\pi^2} K_2(q^2),
\nonumber \\
f_3(q^2)&=& \frac{\LL}{6} + \frac{1}{16\pi^2}
\left[ 2 K_2(q^2) - \frac12 K_0(q^2) \right]
\eea
where
$$ \LL =  \frac{1}{16\pi^2}
\Gamma(\epsilon) \left(\frac{M^2}{4\pi\mu^2}\right)^{-\epsilon} $$
and $K(q^2)$'s are finite functions given explicitly by
\bea
K_0(q^2) &=& \int_0^1 dz\, {\rm ln}\left[1-z(1-z) \frac{q^2}{M^2}\right]
\,=\, -2 + \sigma \, {\rm ln}\left(\frac{\sigma+1}{\sigma-1}\right),
\label{k0q2}\\
K_1(q^2) &=& \int_0^1 dz \frac{-z(1-z) q^2}{M^2 - z(1-z)q^2} 
\,=\, 1-\frac{\sigma^2-1}{2\sigma} \,
{\rm ln}\left(\frac{\sigma+1}{\sigma-1}\right),
\label{k1q2}\\
K_2(q^2) &=& \int_0^1 dz \,z(1-z)\, {\rm ln}\left[1-z(1-z) \frac{q^2}{M^2}
\right] \nonumber \\
&&\ = \  -\frac49 + \frac{\sigma^2}{6} + \frac{\sigma(3-\sigma^2)}{12}
{\rm ln}\left(\frac{\sigma+1}{\sigma-1}\right),
\label{k2q2}\eea
with
\be
\sigma \equiv \left(\frac{4 M^2 - q^2}{-q^2}\right)^{\frac12}\,.
\ee
We should note that all the functions given above
are positive definite for negative $q^2$ and vanish when $q^2=0$.
For $-q^2 \ll M^2$, we have
\bea
K_0(q^2) &=& K_1(q^2) = \frac{1}{6} \, \frac{-q^2}{M^2} + {\cal O}\left(
\frac{q^4}{M^4}\right), \nonumber \\
K_2(q^2) &=& \frac{1}{30} \, \frac{-q^2}{M^2} + {\cal O}\left(
\frac{q^4}{M^4}\right). \nonumber
\eea
In the chiral limit ($-q^2 \gg M^2$), they simplify to
\bea
K_0(q^2) &=& {\rm ln} \frac{-q^2}{M^2} - 2, \nonumber \\
K_1(q^2) &=& 1, \nonumber \\
K_2(q^2) &=& \frac16 {\rm ln} \frac{-q^2}{M^2}- \frac{5}{18} . \nonumber
\eea

In order to go to the $r-$space, we must Fourier-transform
$K_i(q^2)$ and $M^2/(M^2-q^2)\, K_i(q^2)$. This involves an
integration\footnote{Recall that $K_i(q^2)$ does not go
to zero when ${\vec{q}}$ goes to infinity.}
of highly oscillating functions.
Instead of introducing a regulating function which
kills contributions from large ${\vec{q}}^2$ and performing a tricky numerical
calculation, we transform the problem into an integral
of a smooth function with the use of residue calculation.
The point is that we do the parametric integration (of variable $z$)
at the last step.
To see how this work, let us rewrite $K_i(q^2)$,
\bea
K_0(-Q^2) &=& \int_0^1 dx\, {\rm ln} \left(1 + \frac{Q^2}{E^2}\right),
\nonumber \\
K_1(-Q^2) &=& \int_0^1 dx\, \frac{Q^2}{Q^2 + E^2},
\nonumber \\
K_2(-Q^2) &=& \int_0^1 dx\, \frac{1-x^2}{4}\, {\rm ln}
\left(1 + \frac{Q^2}{E^2}\right).
\eea
where $Q= |{\bf q}| = \sqrt{-q^2}$ and $E= E(x)= 2M/\sqrt{1-x^2}$.\footnote{
We have made the change of variable, $x=2 z-1$, to render the expressions
more symmetric.}
First we Fourier-transform algebraically
the integrands of the above integrals and then do the parametric integral
numerically.
The  Fourier transform of $K_1$ becomes an elementary residue calculation
with a pole at $Q= i M$,
\be
K_1(r) = \delta({\bf r}) - \frac{1}{4\pi r} \int_0^1 dx\, E^2\, \e^{- E r}
\ee
while $K_0$ and $K_2$ are somewhat involved due to the logarithmic
function.
We rewrite the logarithmic functions (of $K_0$ and $K_2$)
into a simple pole form by integration by part
\be
\int_0^1 dx\,\left[1,\frac{1-x^2}{4}\right]\,{\rm ln}
\left(1 + \frac{Q^2}{E^2}\right)
\ = \ \int_0^1 dx \,\frac{2 x^2}{1-x^2}\,\left[1,\
\frac14- \frac{x^2}{12}\right]\,\frac{Q^2}{Q^2 + E^2}.
\ee
With the above equation and
\be
\int\frac{d^3{\bf q}}{(2\pi)^3}\, {\rm e}^{i \,{\bf q}\cdot{\bf r}}\,
\frac{Q^2}{Q^2+ E^2}\,\frac{\Lambda^2}{Q^2+\Lambda^2}
= \frac{1}{4\pi r}\,\frac{\Lambda^2}{\Lambda^2- E^2}\,\left(
\Lambda^2 \e^{- \Lambda r} - E^2 \e^{-  E r}\right),
\ee
we obtain the expressions for $K(r,\Lambda)$ defined by
\be
K_i(r,\Lambda) \equiv
\int\frac{d^3{\bf q}}{(2\pi)^3}\, {\rm e}^{i \,{\bf q}\cdot{\bf r}}\,
K_i(-Q^2)\,\frac{\Lambda^2}{Q^2+\Lambda^2},
\ee
\bea
K_0(r,\Lambda) &=& \frac{1}{4\pi r}\, \int_0^1 dx \,
\frac{2 x^2}{1-x^2}\, \frac{\Lambda^2}{\Lambda^2- E^2}\,\left(
\Lambda^2 \e^{- \Lambda r} - E^2 \e^{-  E r}\right),
\nonumber \\
K_2(r,\Lambda) &=& \frac{1}{4\pi r}\, \int_0^1 dx \,
\frac{2 x^2}{1-x^2}\, \left(\frac{1}{4}-\frac{x^2}{12}\right)\,
\frac{\Lambda^2}{\Lambda^2- E^2}\,\left(
\Lambda^2 \e^{- \Lambda r} - E^2 \e^{-  E r}\right).
\eea
Here the parameter $\Lambda$ is introduced to regularize the integrals
near the origin of the configuration space. When $\Lambda= \infty$, we get
the expressions for the $K_i(r)$ needed in the paper,
\be
K_i(r) \equiv \lim_{\Lambda\rightarrow \infty} K_i(r, \Lambda).
\ee
The above integrals are non-singular even near $x=1$,
since $E$ increases so as to make the integrals regular.
However the expressions contain highly singular terms near the
origin of configuration space when $\Lambda$ goes to infinity.
To see this, note
$$
K_0(r) = - \frac{1}{4\pi r}\,
\int_0^1 dx\, \frac{2 x^2}{1-x^2} \, E^2\, \e^{- E r}
+ \lim_{\Lambda\rightarrow \infty}\frac{1}{4\pi r}\,
\Lambda^2 K_0(q^2= \Lambda^2)\, \e^{- \Lambda r},\label{kinfinity}
$$
and
$$
\lim_{\Lambda\rightarrow \infty} \frac{1}{4\pi r}\, \Lambda^2 \e^{-
\Lambda r} = \delta({\bf r}).
$$
Now  $K_0(\Lambda^2)$ goes to infinity logarithmically when $\Lambda^2$
goes to infinity. So, roughly, the second term of (\ref{kinfinity})
behaves in the limit of infinite $\Lambda$ as
$$\lim_{\Lambda^2\rightarrow \infty} {\rm ln}\frac{\Lambda^2}{M^2}\,
\delta({\bf r}). $$
The mathematical reason for this behavior is not hard to see :
For large $Q^2$, both $K_0$ and $K_2$ increase logarithmically with
a gentle slope. Thus they can be viewed as a constant
at large $Q^2$ with their value increasing to infinity.
The constant behavior leads to the $\delta({\bf r})$ function.
This term singular at the origin is of course ``killed" by the short-range
correlation present in nuclear wave functions.

In terms of the function so defined, we can immediately obtain
${\tilde K}_i(r)$ defined by
\be
{\tilde K}_i(r) \equiv
\int\frac{d^3{\bf q}}{(2\pi)^3}\, {\rm e}^{i \,{\bf q}\cdot{\bf r}}\,
K_i(-Q^2)\,\frac{M^2}{Q^2+M^2} \,=\, K_i(r, M).
\ee

In summary we have the expressions for $K_i(r)$ (valid for $r > 0$)
\bea
K_0(r) &=& - \frac{1}{4\pi r}\, \int_0^1 dx \,
\frac{2 x^2}{1-x^2}\, E^2 \e^{-  E r},
\nonumber \\
K_2(r) &=& - \frac{1}{4\pi r}\, \int_0^1 dx \,
\frac{2 x^2}{1-x^2}\, \left(\frac{1}{4}-\frac{x^2}{12}\right)\,
E^2 \e^{-  E r}
\eea
and the expressions for ${\tilde K}_i(r)$ (valid in the whole region)
\bea
{\tilde K}_0(r) &=& \frac{1}{4\pi r}\, \int_0^1 dx \,
\frac{2 x^2}{1-x^2}\, \frac{M^2}{M^2- E^2}\,\left(
M^2 \e^{- M r} - E^2 \e^{-  E r}\right),
\nonumber \\
{\tilde K}_2(r) &=& \frac{1}{4\pi r}\, \int_0^1 dx \,
\frac{2 x^2}{1-x^2}\, \left(\frac14 -\frac{x^2}{12}\right)\,
\frac{M^2}{M^2- E^2}\,\left( M^2 \e^{- M r} - E^2 \e^{-  E r}\right).
\eea
\section*{Appendix C: Integral Identities}
\renewcommand{\theequation}{C.\arabic{equation}}
\setcounter{equation}{0}
\indent\indent In this Appendix, we list some useful identities for
the integrals we need to evaluate.
Consider the following integral without imposing the condition $v^2 = 1$,
$$ I_\alpha(v,M^2) \equiv \int_l \frac{l_\alpha}{v\cdot l \, (l^2-M^2)}, $$
then Lorentz covariance implies that its most general form is
$$ I_\alpha(v,M^2) = v_\alpha I_0(v^2, M^2)\,.  $$
Now multiplying $v^\alpha$ to both sides, we get
$$ \Delta(M^2) = v^2 I_0(v^2,M^2) $$
or
\be
\int_l \frac{l_\alpha}{v\cdot l \, (l^2-M^2)}
= \frac{v_\alpha}{v^2} \Delta(M^2)\,.
\ee
Using this expression, we obtain the following identities by
successive differentiation with respect to $v$,
\bea
\int_l \frac{l_\alpha l_\beta}{(v\cdot l)^2 \, (l^2-M^2)} &=&
-\left(\frac{g_{\alpha\beta}}{v^2} -2 \frac{v_\alpha v_\beta}{(v^2)^2}
\right)  \Delta(M^2) \\
\int_l \frac{l_\alpha l_\beta l_\mu}{(v\cdot l)^3 \, (l^2-M^2)} &=&
\left[ 4 \frac{v_\alpha v_\beta v_\mu}{(v^2)^3} -
\frac{v_\mu g_{\alpha \beta}
+ v_\alpha g_{\beta\mu} + v_\beta g_{\alpha\mu}}{(v^2)^2}\right]
\Delta(M^2)
\eea
and so on.
\section*{Appendix D: Functions $h(v\cdot k)$'s}
\renewcommand{\theequation}{D.\arabic{equation}}
\setcounter{equation}{0}
\indent \indent
In this Appendix, we define -- and give explicit forms of -- the function
$h(v\cdot k)$ that figures in the self-energy for the off-shell nucleon
discussed in section (4.3). The basic one is $h_0(v\cdot k)$,
\bea
h_0(v\cdot k) &=& \int_l \frac{1}{v\cdot(l+k)\,\,(l^2-M^2)}
\nonumber \\
&=& 2 y \LL + \frac{1}{8\pi^2}\left[
\pi M_* + 2 y - 2 M_* \sin^{-1}\left(\frac{y}{M}\right)
\right],\ \ \ \ \mbox{for}\ \ \ |y| \leq M,
\\
&=& 2 y \LL + \frac{1}{8\pi^2}\left[
2 y - 2 {\tilde M} \sinh^{-1}\left(\frac{y}{M}\right)
- \theta(y-M) 2i\pi {\tilde M}\right], \ \mbox{for}\ \ |y| \geq M\nonumber\\
\eea
where $y\equiv v\cdot k$, $M_* \equiv \sqrt{M^2-y^2}$, ${\tilde M} \equiv
\sqrt{y^2-M^2}$ and $-\frac{\pi}{2} \leq \sin^{-1}(x) \leq \frac{\pi}{2}$.
We observe that the imaginary part appears only when $v\cdot k \geq M$.
The even part of $h_0$ has a very simple form
\be
h_0^S(y) \equiv \frac{h_0(y) + h_0(-y)}{2}
= -\frac{1}{8\pi}\sqrt{M^2-y^2}.
\ee
For special values of $y$, we have $h_0(0)= -\frac{M}{8\pi}$, $h_0'(0)= 2 \LL$,
$h_0(M) = - h_0(-M) = \frac{2M}{1-2\epsilon} \LL$.
The finite function ${\overline h_0}(y)$ is defined by
\be \overline{h_0}(y) = h_0(y) - 2 y \LL.\ee

We now examine the function $h(v\cdot k)$ defined by
\be
\int_l \frac{l_\alpha \, l_\beta}{v\cdot(l+k)\,(l^2-M^2)}
\equiv g_{\alpha\beta} h(v\cdot k) + v_\alpha v_\beta (\cdots).
\ee
If we multiply the above equation by $g^{\alpha\beta}$ and $v^\beta$,
we obtain the following identity
\be
h(y) = \frac{1}{d-1} \left[ y\, \Delta(M^2) + (M^2-y^2) h_0(y)\right].
\label{hy}\ee
Note that the even part has a very simple form,
\be
h^S(y) \equiv \frac{h(y) + h(-y)}{2} = -\frac{1}{24\pi} \left(
M^2-y^2\right)^\frac32.
\ee
Let us define $h_3(v\cdot k, v\cdot q)$
$$
\int_l \frac{l_\alpha l_\beta}{v\cdot(l+k-q)\, v\cdot(l+k)\, (l^2-M^2)}
= g_{\alpha\beta} h_3(v\cdot k, v\cdot q) + v_\alpha v_\beta (\cdots)
$$ or
\be
h_3(v\cdot k, v\cdot q) =\frac{1}{v\cdot q}\left[h(v\cdot k
-v\cdot q)- h(v\cdot k)\right].
\ee
When the nucleon is on-shell, that is, $v\cdot k=v\cdot q=0$, then
$h_3$ becomes $h_3(0,0) = - \Delta(M^2)$. More generally, for small
momentum, we have
\be
h_3(v\cdot k, v\cdot q) = -\Delta(M^2) - \frac{M}{16\pi}(2v\cdot k
-v\cdot q) + \frac{2}{3}\LL\left[3(v\cdot k)^2 -
3v\cdot k \, v\cdot q + (v\cdot q)^2\right] + \cdots
\ee
where the ellipsis denotes finite and higher momentum terms.
Finally consider $h_4(v\cdot q)$ defined by
\be
\int_l \frac{l_\alpha \, l_\beta}{(v\cdot l)^2\,v\cdot(l-q)\,(l^2-M^2)}
\equiv g_{\alpha\beta} h_4(v\cdot q) + v_\alpha v_\beta (\cdots).
\ee
The $h_4$ is a somewhat complicated function,
\bea
h_4(y) &=& \frac{h(-y) - h(0)}{y^2} + \frac{\Delta(M^2)}{y}
\nonumber \\
&=& \frac{M}{16\pi} + \frac23 v\cdot q\, \LL + \cdots
\eea
with $h(y)$ given by (\ref{hy}) and the ellipsis again denotes
finite and higher momentum terms.

\section*{Appendix E: Integrals for Two-Pion Exchange Currents}
\renewcommand{\theequation}{E.\arabic{equation}}
\setcounter{equation}{0}
\indent \indent
Consider the integrals of the form
$$\int_l \frac{l\mbox{'s}}{v\cdot(l+k)\, (l^2-M^2)\,\left[(l+q)^2-M^2
\right]} $$
which figure in two-pion exchange currents.
For most of the cases, we do not need the terms proportional to $v^\mu$
as they appear multiplied by the spin operator $S_\mu$ and vanish.
To utilize this, we assume that the
spin operator is multiplied to the numerator.
Now we have
\bea
\int_l \frac{(l+q)^\alpha\,l^\beta}{
v\cdot(l+k)\, (l^2-M^2)\,\left[(l+q)^2-M^2 \right]}
&=& \left[ g^{\alpha\beta} + 2 q^\alpha q^\beta \frac{\partial}{
\partial q^2}\right] B_0(k,q),
\\
\int_l \frac{(l+q)^\alpha\, l^\beta\, (2l+q)^\mu}{
v\cdot(l+k)\, (l^2-M^2)\,\left[(l+q)^2-M^2 \right]}
&=& \left( g^{\alpha\beta} + 2 q^\alpha q^\beta \frac{\partial}{
\partial q^2}\right) \left[q^\mu B_1(k,q) + v^\mu B_2(k,q)\right]
\nonumber \\
+ \left(q^\alpha\, g^{\beta\mu} + q^\beta g^{\alpha\mu}\right) B_1(k,q)
\!&+&\! \left(q^\alpha g^{\beta\mu}-q^\beta g^{\alpha\mu}\right) B_0(k,q)
\eea
where we have neglected terms proportional to $v^\alpha$ or $v^\beta$.
After some algebra, we can get the following relations,
\bea
&& q^2 B_1(k,q) + v\cdot q B_2(k,q) = h(v\cdot k) - h(v\cdot k - v\cdot q),
\nonumber \\
&& B_2(k,q) = f_1(q^2) + v\cdot q \left[ B_0(k,q)-B_1(k,q)\right]
- 2 v\cdot k B_0(k,q).
\nonumber \eea
When $v\cdot k = v\cdot q = 0$, they become elementary functions,
\bea
B_0(q^2) &=& -\frac{1}{16\pi} \int_0^1 dz \sqrt{M^2 - z(1-z)q^2},
\nonumber \\
B_1(q^2) &=& 0, \nonumber \\
B_2(q^2) &=& f_1(q^2)
\eea
For small but nonzero momentum, they become
\bea
B_0(k,q) &=& -\frac{M}{16\pi} + \left(v\cdot k - \frac12 v\cdot q\right)
\LL + \cdots,
\nonumber \\
B_1(k,q) &=& \frac{v\cdot q}{6} \LL + \cdots,
\nonumber \\
B_2(k,q) &=& \Delta(M^2) + (2 v\cdot k - v\cdot q) \frac{M}{16\pi}
- \left[2(v\cdot k)^2 - 2 v\cdot k \, v\cdot q + \frac23 (v\cdot q)^2
+ \, \frac{q^2}{6}\right] \LL
\nonumber \\
&& + \, \cdots.
\eea

For two pion exchange graphs, we need to evaluate
\be
I_{\nu, \alpha\beta} \equiv
\int_l \frac{(l+q)_\nu\,l_\alpha\,l_\beta}{v\cdot l\, v\cdot(l+q)\,
(l^2-M^2)\,\left[(l+q)^2-M^2\right]}.
\nonumber \ee
In evaluating this function, we neglect terms proportional to
$v_\nu$, $v_\alpha$ or $v_\beta$ because they vanish when multiplied
by $S^\nu S^\alpha S^\beta$. With the parametrization explained in
the text, we have, in the limit of $v\cdot q=0$,
\be
I_{\nu,\alpha\beta}(q) = -\frac{1}{32 \pi^2} (-q_\nu g_{\alpha\beta}
+ q_\alpha g_{\nu\beta} + q_\beta g_{\nu\alpha}) \left[ K_0(q^2) -16\pi^2 \LL
\right] - \frac{1}{16\pi^2} \frac{q_\nu q_\alpha q_\beta}{q^2} K_1(q^2),
\ee
where $K_1(q^2)$ is defined at (\ref{k1q2}),
\be
K_1(q^2) = \int_0^1 dz \frac{-z(1-z) q^2}{M^2 - z(1-z) q^2}.
\ee

For completeness, we also list results for vector currents for which
we need the following integrals
\be
\int_l \frac{(l+q)^\alpha \, l^\beta \,(2l+q)^\mu}{
v\cdot l\,v\cdot(l+q) \,(l^2-M^2)\,\left[(l+q)^2-M^2\right]}.
\nonumber \ee
We first look at its low momentum behavior,
\be
- v^\mu \frac{M}{8\pi} g^{\alpha\beta}
-\left(q^\alpha \, g^{\beta\mu} - q^\beta \,g^{\alpha\mu}\right) \LL
+ {\cal O}\left(q^2\right).
\ee
In the limit of $v\cdot q=0$, we have
\bea
&& \hspace{-1.2cm}
\int_l \frac{(l+q)^\alpha \, l^\beta \,(2l+q)^\mu}{
(v\cdot l)^2\,(l^2-M^2)\,\left[(l+q)^2-M^2\right]}
\nonumber \\
&=&\hspace{-0.3cm}  -\frac{v^\mu}{8\pi} \left(g^{\alpha\beta} + 2 q^\alpha
q^\beta \frac{\partial}{\partial q^2}\right) \int_0^1 dz
\sqrt{M^2- z(1-z)q^2}
 -  \left(q^\alpha\,g^{\beta\mu} - q^\beta \,g^{\alpha\mu}\right)
f_0(q^2).
\eea
Again we dropped terms proportional to $v^\alpha$ or $v^\beta$.

\section*{Appendix F: Three-point Vertices (Figure 4)}
\renewcommand{\theequation}{F.\arabic{equation}}
\setcounter{equation}{0}
\indent \indent
In this section as well as in the next, we classify graphs into
Class A, Class V and Class AV. The graphs in Class V appear
only with the vector current while the graphs in Class AV appear both for
the vector and axial-vector currents.
The graphs in Class A involving the axial-vector current do not figure in
three-point vertices. We define two operators for the graphs in Class AV:
$T_1^\mu= v^\mu$ and $T_2 = 2 \gA S^\mu$. We write the expressions only
for the axial-vector
current for graphs in Class AV. The expressions for the
vector current is obtained by interchanging $T_1^\mu$ and $T_2^\mu$.
Fig.4$c$ and 4$d$ vanish because they are proportional to $v\cdot S$.
\\ \underline{Class AV}\footnote{The figure label $a,b,c...$ is given in
the parenthesis.}
\bea
\GANN(a) &=& \frac{\tau_a}{2}\frac{T_2^\mu}{F^2} \Delta(M^2),
\\
\GANN(b) &=& \frac{\tau_a}{2}\frac{\gA^2}{F^2} \, S_\alpha T_2^\mu S_\alpha
\,h_3(v\cdot k, v\cdot q), \\
i\GpiNN(a) &=& \frac{\gA}{3F^3} \tau_a \,q\cdot S\, \Delta(M^2),
\\
i\GpiNN(b) &=& \frac{d-3}{4}\tau_a \,q\cdot S\,\frac{\gA^3}{F^3}
\,h_3(v\cdot k, v\cdot q),
\eea

\underline{Class V}
\bea
\GVNN(e) &=& -\frac{\tau_a}{2F^2} \left[ v^\mu f_1(q^2) + v\cdot q\, q^\mu
f_3(q^2)\right], \\
\GVNN(f) &=& 2 \tau_a \frac{\gA^2}{F^2} \left\{ \left[S\cdot S +
2 (q\cdot S)^2 \frac{\partial}{\partial q^2}\right]\left[ q^\mu B_1(k,q)
+v^\mu B_2(k,q)\right] \right. \nonumber \\
&& \left. \ \ \ \ \ \ \ \
+\left\{q\cdot S, S^\mu\right\}B_1(k,q) + \left[ q\cdot S, S^\mu\right]
B_0(k,q)\right\}
\eea
\bea
\GVpipi(g) &=& i \eabc (q_c-q_b)^\mu \frac53 \frac{\Delta(M^2)}{F^2},
\\
\GVpipi(h) &=& i\eabc \frac{1}{F^2}\left[
-(q_c-q_b)^\mu f_1(q_a^2) + q_a^\mu\,(q_c^2-q_b^2) f_3(q_a^2)\right].
\eea
Here the index $a$ labels the isospin of the photon with four-momentum
$q_a$, the indices $b$ and $c$ the isospin of the pions
with their momenta $q_b$ and $q_c$ and the momentum conservation
is $q_a+q_b+q_c=0$.

\section*{Appendix G: Four-point Vertices (Figure 5)}
\renewcommand{\theequation}{G.\arabic{equation}}
\setcounter{equation}{0}
\indent \indent
Here we define $q^\mu = q_a^\mu + q_b^\mu$. For other notations,
see Appendix F.
Figures 5$(i)-(n)$ vanish because they are proportional to $v\cdot S$.
Here we restrict ourselves to the case of
on-shell nucleons, $v\cdot k=v\cdot (q_a+q_b)=0$.
\\
\underline{Class A}
\bea
\GpiA(a) &=& \frac{1}{2F^3} \eabc\tau_c\, v^\mu f_1(q^2),
\label{A1} \\
\GpiA(b) &=& -i\dab (2q_a+6q_b)^\mu\frac{2 \gA^2}{3 F^3}
\left[ S\cdot S + 2 (q\cdot S)^2 \frac{\partial}{\partial q^2}\right]
B_0(q^2)\nonumber \\
&&- \eabc\tau_c\, \frac{2 \gA^2}{F^3}\left\{
v^\mu \left[S\cdot S+2 (q\cdot S)^2 \frac{\partial}{\partial q^2}\right]
f_1(q^2) + \left[ q\cdot S, S^\mu\right] B_0(q^2)\right\},\nonumber\\
\label{A2} \eea
\underline{Class AV}
\bea
\GpiA(c) &=& -\frac{5}{12} \eabc\tau_c \frac{T_1^\mu}{F^3}
\Delta(M^2), \label{AV1} \\
\GpiA(d) &=& \frac{\gA^2}{2F^3} \eabc\tau_c\,
S_\alpha T_1^\mu S^\alpha \, \Delta(M^2),
\label{AV2}\\
\GpiA(e+f) &=& \frac{T_1^\mu}{4F^3} \left\{
4i\dab\,v\cdot q_a \, h_0^S(v\cdot q_a) + \eabc\tau_c \left[
\Delta(M^2) - 2 v\cdot q_a h_0^A(v\cdot q_a)\right]\right\},\nonumber\\
\label{AV3} \\
\GpiA(g+h) &=& (-3i\dab + \eabc\tau_c) \frac{\gA^3}{2F^3}
S_\alpha\, q_b\cdot S \, T_2^\mu \, S^\alpha \, h_4(v\cdot q_a) \nonumber \\
&&+ (-3i\dab - \eabc\tau_c) \frac{\gA^3}{2F^3}
S_\alpha\, T_2^\mu\, q_b\cdot S \, S^\alpha \, h_4(-v\cdot q_a),
\label{AV4} \eea
\underline{Class V}
\bea
\GpiV(o+p+q) &=& \frac{\gA}{2F^3} v\cdot q_a \left\{
(-2i\dab -\eabc \tau_c)\int_l
\frac{l\cdot S \, (2l+q_a)^\mu}{v\cdot l\,
(l^2-M^2)\,\left[(l+q_a)^2-M^2\right]}\right.\nonumber \\
&&\left.  \ \ \ \ \ \ - (2i\dab - \eabc\tau_c) \int_l
\frac{l\cdot S \, (2l-q_a)^\mu}{v\cdot l\,
(l^2-M^2)\,\left[(l-q_a)^2-M^2\right]}\right\},
\label{V2} \\
\GpiV(r) &=& -2i\dab \frac{\gA^3}{F^3} \int_l \frac{
(l+q_a)\cdot S \, q_b\cdot S \, l\cdot S (2l+q_a)^\mu}{
v\cdot(l+q_a)\, v\cdot l \, (l^2-M^2)\, \left[(l+q_a)^2-M^2\right]}
\label{V3} \eea
\indent where
\bea
B_0(q^2) &=& -\frac{1}{16\pi} \int^1_0 dz\, \sqrt{M^2-z(1-z) q^2}, \\
h_0^S(v\cdot q) &=& \frac12 \left[h_0(v\cdot q) + h_0(-v\cdot q)\right]
\, = \, -\frac{1}{8\pi} \sqrt{M^2- (v\cdot q)^2},\\
h_0^A(v\cdot q) &=& \frac12 \left[h_0(v\cdot q) - h_0(-v\cdot q)\right].
\eea
Now we study {\it low-momentum expansion for on-shell nucleons with
$v\cdot k= v\cdot (q_a+q_b)=0$}. To second order in external momentum;
\\
\underline{Class A}
\bea
\GpiA(a)
&=& \frac{1}{2F^3} \eabc\tau_c\, v^\mu \left[\Delta(M^2)
-\frac{\LL}{6} q^2 + \cdots\right],
\nonumber  \\
\GpiA(b)
&=& -i\dab (q_a+3q_b)^\mu\frac{\gA^2}{F^3} M' \nonumber \\
&&- \eabc\tau_c\, \frac{2 \gA^2}{F^3}\left\{
v^\mu \,S\cdot S \Delta(M^2)- v^\mu \frac{\LL}{6}
\left[S\cdot S\, q^2 +2 (q\cdot S)^2\right]
- \left[q\cdot S, S^\mu\right] M'\right\}
\nonumber \\
&& + \  \cdots \, .
\nonumber  \eea
\underline{Class AV}
\bea
\GpiA(c) &=& -\frac{5}{12} \eabc\tau_c \frac{T_1^\mu}{F^3}
\Delta(M^2), \nonumber   \\
\GpiA(d) &=& \frac{\gA^2}{2F^3} \eabc\tau_c\,
S_\alpha T_1^\mu S^\alpha \, \Delta(M^2),
\nonumber  \\
\GpiA(e+f)
&=& \frac{T_1^\mu}{4F^3} \left\{
-8i\dab\,v\cdot q_a \, M' + \eabc\tau_c \left[
\Delta(M^2) - 4 (v\cdot q_a)^2 \,\LL \right]\right\}\cdots,
\nonumber   \\
\GpiA(g+h)
&=& (-3i\dab + \eabc\tau_c) \frac{\gA^3}{2F^3}
S_\alpha\, q_b\cdot S \, T_2^\mu \, S^\alpha \, \left[M'
+ \frac23 \, v\cdot q_a \, \LL \right]
\nonumber \\
&&+ (-3i\dab - \eabc\tau_c) \frac{\gA^3}{2F^2}
S_\alpha\, T_2^\mu\, q_b\cdot S \, S^\alpha \, \left[M'
- \frac23 \, v\cdot q_a \, \LL\right]
+ \cdots .
\nonumber  \eea
\underline{Class V}
\bea
\GpiV(o+p+q)
&=& \frac{\gA}{F^3}\,v\cdot q_a\left[4i\dab M' S^\mu
+ \eabc\tau_c (S^\mu v\cdot q_a + v^\mu q_a\cdot S) \LL\right]
+ \cdots,
\nonumber  \\
\GpiV(r)
&=& -2i\dab \frac{\gA^3}{F^3} \left\{
-2 v^\mu M' S_\alpha q_b \cdot S \, S^\alpha
- (q_a \cdot S \, q_b\cdot S\, S^\mu- S^\mu\, q_b\cdot S\, q_a\cdot S)\LL
\right\}
\nonumber \\
&& + \cdots.
\nonumber  \eea
We have used the notations
$M' = \frac{M}{16\pi}$ and $\LL= \frac{1}{16\pi^2}
\left(\frac{M^2}{4\pi\mu^2}\right)^{-\epsilon}\Gamma(\epsilon)$.
\section*{Appendix H: Fermi-Gas Model for Two-Body Axial-Charge Operator}
\renewcommand{\theequation}{H.\arabic{equation}}
\setcounter{equation}{0}
\indent \indent
Let $|F\rangle$ be the ground state of Fermi-gas model whose
fermi-momentum is $\pF$ and $|ph\rangle= b_p^\dagger b_h |F\rangle$ be the
one-particle (labeled by $p$) one-hole (labeled by $h$)
excited state, where $b_\alpha(b_\alpha^\dagger)$ is the
annihilation(creation) operator of a fermion state characterized by $\alpha$.
Consider the matrix element
$\langle ph | {\cal M} |F\rangle$ (or its {\it effective
one body operator} ${\cal M}_{\mbox{eff}}$),
\be
\langle ph|{\cal M}| F\rangle = \langle p| {\cal M}_{\mbox{eff}} | h\rangle
= \sum_{\beta \in F} \frac{1}{g_\beta} \langle p,\beta | {\cal M} | h
,\beta\rangle
\ee
where $g_\beta$ is defined by
$\left\{ b_\alpha, b_\beta^\dagger\right\} = g_\beta \delta_{\alpha,\beta}$.
In computing this, it is convenient to define the antisymmetrized wave
function $|\alpha,\beta\rangle$ in terms of the simple
two-particle state $|\alpha,\beta)$
\be
|\alpha,\beta\rangle= \frac{|\alpha,\beta ) - |\beta,\alpha)}{\sqrt{2}},
\nonumber \ee
so the matrix element $\langle p,\beta |{\cal M}|h,\beta\rangle$
is of the form
\be
\langle p, \beta |{\cal M}| h, \beta\rangle = (p, \beta |{\cal M}| h,\beta)
-(p, \beta |{\cal M}| \beta, h).
\nonumber \ee
The first term is the Hartree term and the second the Fock term.
Rewriting $|\alpha\rangle$ as $|\bfp_\alpha m_\alpha t_\alpha\rangle$
where $\bfp_\alpha$ is the momentum of the state labeled by $\alpha$ and
$m_\alpha$ ($t_\alpha$) the third component of the spin
(isospin) of the state $\alpha$, we may write the axial charge operator as
\be
(1', 2'| {\cal M} |1,2) = (t_1'm_1' , t_2'm_2'| \left[
{\cal T}^{(1)} \phi_1(q) + {\cal T}^{(2)} \phi_2(q)\right] | t_1 m_1, t_2 m_2)
\nonumber \ee
where
${\cal T}^{(1)}= i\,\tauvec_1\times\tauvec_2 ({\bfsigma}_1+{\bfsigma}_2)
\cdot {\bf q}$,
${\cal T}^{(2)}= i\,(\tauvec_1+\tauvec_2) {\bfsigma}_1\times{\bfsigma}_2
\cdot {\bf q}$
and $\bfq=\bfp_2'-\bfp_2=\bfp_1-\bfp_1'$, $q\equiv |\bfq|$.

It is trivial to see that the Hartree term must vanish. Thus we are to
calculate
the Fock term. First we shall show that the matrix element of ${\cal T}^{(1)}$
is equal to that of ${\cal T}^{(2)}$ when summed over spin and isospin
of occupied states. Doing the sum, we get
\be
\sum_{m_\beta} \sum_{t_\beta} (m_p t_p,m_\beta t_\beta| {\cal T}^{(1)}|
m_\beta t_\beta, m_h t_h)
= -4 \, \langle m_p t_p | \tauvec \otimes {\bfsigma}|m_h t_h\rangle
\cdot {\bf q}.
\ee
For ${\cal T}^{(2)}$, we simply interchange the spin and isospin operators
and get the same result. It then follows that
the effective one-body operator of the axial-charge operator becomes
\be
\langle p | {\cal M}_{\mbox{eff}} |h \rangle
= 4 \tauvec_{ph} \bfsigma_{ph}
 \, \int_{|\bfp| < \pF} \frac{d\bfp}{(2\pi)^3} \,
(\bfp-\bfp_h)\, \phi(|\bfp-\bfp_h|)
\ee
where
$\phi(q)=\phi_1(q) + \phi_2(q)$, $\pF$ stands for the fermi-momentum
of the ground state $|F\rangle$ and
$\tauvec_{ph} ({\bfsigma}_{ph})$ is
$\langle t_p | \tauvec|t_h\rangle$ ($\langle m_p | {\bfsigma} |m_h\rangle$).
This form is particularly useful when the particle is on the
fermi surface $|\bfp_h|=\pF$,
\be
\int_{|\bfp| < \pF} \frac{d\bfp}{(2\pi)^3} \,
(\bfp-\bfp_h)\, \phi(|\bfp-\bfp_h|)
= -\frac{\hat{\bfp}_h}{4\pi^2} 8 \pF^4\int_{0}^1 dx\,
(x^3-x^5) \, \phi(2 x \pF).
\ee
In order to give a meaning to this expression, we have to account for
short-range correlations. Otherwise we can get erroneous results.
For instance, a constant in momentum space (say, $\phi (q)=
{\rm constant}$) gives a contribution whereas it should be suppressed in
reality. One way to assure a correct behavior at short-distance
is to subtract the constant as one does for the Lorentz-Lorenz
effect in $\pi$-nuclear scattering. However this procedure is not always
practicable if one is dealing with non-polynomial terms.
It is therefore preferable to go to
coordinate space by Fourier-transform. For this, define $f(r)$ by
\be
f(r) = \int \frac{d \bfq}{(2\pi)^3}\,\e^{i\bfq\cdot\bfr}
\,\phi(q)\,.
\ee
Using
\bea
\int_{|\bfp| < \pF} \frac{d\bfp}{(2\pi)^3}\, \e^{i\bfp\cdot\bfr}
&=& \frac{\pF^2}{2\pi^2 r} j_1(\pF r),\ \
j_1(x)= \frac{\sin x}{x^2}- \frac{\cos x}{x}
\nonumber \\
\mbox{and} \ \ \ \ \ \ \ \ \ \ \ \ \ \ \ \ \ \
\int d\Omega\, {\hat r} \e^{i\bfp\cdot\bfr}&=& i\,4\pi \hat{\bfp} j_1(p r),
\nonumber \eea
we obtain
\be
\langle ph | {\cal M} | F\rangle
= \tauvec_{ph} \,\bfsigma_{ph}\cdot\hat{\bfp}_h\,
\frac{8 \pF^2}{\pi} \int_0^\infty d r
\, r\, j_1(\pF r) j_1(p_h r)\, \frac{d}{dr} f(r).
\ee
Introducing a correlation function $\hat{g}(r,d)$ where $d$ is
a parameter of $\hat{g}$,  we get the final expression
\be
\langle ph | {\cal M} |F\rangle
= \tauvec_{ph} \,\bfsigma_{ph}\cdot\hat{\bfp}_h\,
\frac{8 \pF^2}{\pi} \int_d^\infty d r
\, r\, j_1(\pF r) j_1(p_h r)\, \frac{d}{dr}f(r).
\ee
In the numerical results discussed in the next, we have used the
simplest correlation function, $\hat{g}(r,d)= \theta(r-d)$.

\section*{Appendix I: The Role of Vector Mesons}
\renewcommand{\theequation}{I.\arabic{equation}}
\setcounter{equation}{0}
\indent \indent
In this Appendix, we describe briefly the role that vector mesons
play in the axial-charge transitions in chiral perturbation theory.
We will in particular establish that vector mesons can contribute
only at ${\cal O} (\frac{1}{m^2})$ and hence their contributions
are suppressed to the chiral order
we are concerned with. For simplicity, we shall consider the vector
field $V_\mu$ only. The axial vector field $a_{1\mu}$ could also be
included but we shall leave it out since it plays even less significant
role in our case. Let
\be
V_\mu = t_a V_\mu^a, \ \ \ \ \ \ \ \
{\rm Tr} (t_a t_b) = \frac12 \delta_{ab}
\ee
denote the spin-1 field. The index $a$ and $b$ are $(1,2,3)$ for SU(2) with
${\vec t} = \frac{{\vec \tau}}{2}$, and $(0,1,2,3)$ for
U(2) with $t_0= \frac12$. The $a=(1,2,3)$ components correspond to
the $\rho$ mesons and $a=0$ to the $\omega$ meson. We write the relevant part
of the Lagrangian as\footnote{We have not included the axial-vector field
$a_1$, although it is not difficult to do so. For the axial-charge process
we are considering the $a_1$ field does not play an important role. For the
Gamow-Teller operator, however, the axial field may not be ignorable.}
\bea
{\cal L} &=& {\bar N} \left[\gamma^\mu \left( i\del_\mu +  g V_\mu + g_A
\gamma_5\, i\Delta_\mu\right) - m\right] N
+ \frac{F^2}{2} \langle  i\Delta_\mu\,
i\Delta^\mu \rangle  + \frac14 M^2 F^2 \langle  \Sigma\rangle
\nonumber \\
&+& \frac{1}{2} M_V^2 \langle  \left(V_\mu -\frac{i}{g} \Gamma_\mu\right)^2
\rangle
-\frac{1}{4} \langle V_{\mu\nu} V^{\mu\nu}\rangle
+{\cal L}_{an} \label{lvector}
\eea
where
\be
V_{\mu\nu} = \del_\mu V_\nu - \del_\nu V_\mu - i g \left[ V_\mu,V_\nu\right],
\ee
and $\Gamma_\mu$ ($\Delta_\mu$) were given in Section 3 and explicitly take
the form
\bea
i\Gamma_\mu &=& {\vec t} \cdot \left[ {\vec {\cal V}}_\mu
+ \frac{1}{F} {\vec \pi}\times {\vec {\cal A}}_\mu
- \frac{1}{2 F^2} {\vec \pi}\times \del_\mu {\vec \pi} + \cdots \right],
\nonumber \\
i\Delta_\mu &=& {\vec t} \cdot \left[ {\vec {\cal A}}_\mu
+ \frac{1}{F} {\vec \pi}\times {\vec {\cal V}}_\mu
- \frac{1}{F} \del_\mu {\vec \pi} + \cdots \right]
\eea
where ${\cal V}_\mu$ (${\cal A}_\mu$) is the external vector (axial-vector)
field and the ellipsis denotes terms involving more than three fields.
Here ${\cal L}_{an}$ is an ``anomalous parity" piece involving the totally
antisymmetric $\epsilon$ tensor which we do not explicit
here as it does not contribute. Also four-fermion interaction terms do not
figure in the discussion.
In (\ref{lvector}), the constants $g$ and $M_V$ can be identified as
the $VNN$ coupling constant and the mass of the $V$ meson respectively.
We are using the short-hand notation
\be
\langle  X \rangle  \equiv 2\, {\rm Tr}(X)
\ee
for any $X$. This convention is convenient due to the normalization
of $t_a$, $\langle X Y \rangle = X_a Y_a$ for any $X= t_a X_a$ and
$Y= t_a Y_a$.

Before proceeding, let us note a few characteristics of this Lagrangian:
\bitem
\item It is vector gauge-invariant (or hidden gauge invariant)
provided that $g V_\mu$ transforms as $i\Gamma_\mu$ does,
$V_\mu \rightarrow U V_\mu U^\dagger -\frac{i}{g} \del_\mu U \cdot U^\dagger$.
It is also invariant under (global) chiral transformation apart from the pion
mass term.
\item It has vector-meson dominance.
\item When $M_V$ goes to infinity, we recover our previous chiral Lagrangian
involving only $\pi$'s and nucleons $N$.
\item There is a $V\gamma$ mixing but the mixing is trivial
in the sense that the photon field appears only as an external
(non-propagating)
field.
\eitem

For the reasons spelled out in the main text, we wish to transform the
Lagrangian to a form appropriate for heavy fermion formalism. Including
the ``$1/m$" terms, we have
\bea
{\cal L} &=& {\bar B} \left( i v\cdot D + 2 g_A \, S\cdot i\Delta\right) B
+ \frac{F^2}{2} \langle  i\Delta_\mu\,
i\Delta^\mu \rangle  + \frac14 M^2 F^2 \langle  \Sigma\rangle
\nonumber \\
&+& \frac{1}{2m} {\bar B} \left( - D^2 + (v\cdot D)^2 +
\left[S^\mu,S^\nu\right]
\left[D_\mu, D_\nu\right] - g_A^2 (v\cdot i\Delta)^2 - 2 i g_A \left\{v\cdot
i\Delta , S\cdot D\right\} \right) B
\nonumber \\
&+& \frac{1}{2} M_V^2 \langle  \left(V_\mu -\frac{i}{g}\Gamma_\mu\right)^2
\rangle
-\frac{1}{4} \langle V_{\mu\nu} V^{\mu\nu}\rangle +{\cal L}_{an}\label{I6}
\eea
where $D_\mu = \del_\mu - i g V_\mu$.
Now let us calculate the tree-order contribution of the vector mesons to
the two-body axial charge operator. Three types of graphs contribute.
The relevant graphs are given in Figure 11.

First we find that the graph $(c)$ does not contribute. To see this, note that
G-parity does not allow the couplings $\rho\rho {\cal A}_\mu$ and
$\omega\omega {\cal A}_\mu$. The coupling for ${\cal A}_\mu\rho\omega$ is
of the form $\epsilon^{\mu\nu\alpha\beta}\omega_{\nu\alpha}\rho_\beta$ coming
from the anomalous-parity term ${\cal L}_{an}$ of (\ref{I6}). In the figure
$(c)$, each vector meson brings in $v_\mu$ as one can see in (\ref{I6}), so
that
we have $\epsilon^{\mu\nu\alpha\beta}v_\alpha v_\beta=0$.

Working out the graphs $(a)$ and $(b)$ \footnote{In the spin-1 propagator
\be
D_{ab}^{\mu\nu}(q) =
\frac{\delta_{ab}}{q^2- M_V^2} \left(
-g^{\mu\nu}+\frac{q^\mu q^\nu}{M_V^2}\right).\nonumber
\ee
the term proportional to $\frac{1}{M_V^2} q^\mu q^\nu$
is a correction term of ${\cal O} \left(\frac{Q^2}{M_V^2}\right)$ relative
to the leading term ($\propto g^{\mu\nu}$). It is further suppressed
in the vector-meson exchange between two nucleons since
the $VNN$ vertex function is proportional to
$v^\mu$ and $v\cdot q = {\cal O}\left(\frac{Q^2}{m}\right)$.
So effectively this term is of order $\frac{Q^3}{m M_V^2}$  and hence can
be dropped.} we get
\bea
{\vec A}^\mu(1) &=& i \tauvec_1\times\tauvec_2 \frac{g_A}{2 \Fp^2}
\frac{M_V^2}{M_V^2-q_1^2} \frac{1}{\Mp^2-q_2^2} \left(q_2\cdot S_2 -
\frac{v\cdot q_2}{m_N} S_2\cdot P_2\right)
\nonumber \\
&&\otimes\ \left(v_1^\mu + \frac{1}{m_N} \left[ S_1^\mu, q_1\cdot S_1\right]
- \frac{v_1\cdot q_1}{M_V^2} q_1^\mu\right) \ + \ (1\leftrightarrow 2)
\\
{\vec A}^\mu(2) &=& (\tauvec_1 +\tauvec_2) \frac{g_A}{2 \Fp^2}
\frac{M_V^2}{M_V^2-q_2^2}
\frac{v^\mu}{4m_N}\left(v_2 \cdot S_2 + \frac{1}{m_N} \left[S_1\cdot S_2,
q_2\cdot S_2\right] - \frac{v_2\cdot q_2}{M_V^2} q_2\cdot S_2\right)
\nonumber \\
&&+ \ (1\leftrightarrow 2)
\eea
where we have used the KSRF relation $M_V^2 = 2 g^2 \Fp^2$ and defined
\bea
P_i^\mu &=& \frac{1}{2} (p_i + p_i')^\mu,
\nonumber \\
v_i^\mu &=& v^\mu + \frac{1}{m_N} \left( P_i^\mu - v^\mu \, v\cdot P_i\right)
\eea
and $q_i = p_i' - p_i$, $i=1,2$.
Now noting that $v\cdot q \simeq v_i \cdot q_j =
{\cal O}\left(\frac{Q^2}{m_N}\right)$
and $S_i^0 = {\cal O}\left(\frac{Q}{m_N}\right)$, we have (setting
$q_2 = - q_1 \equiv q$)
\bea
{\vec A}^0(1) &=& i \tauvec_1\times\tauvec_2 \frac{g_A}{2 \Fp^2}
\frac{M_V^2}{M_V^2-\Mp^2} \left(\frac{1}{\Mp^2-q^2} -\frac{1}{M_V^2-q^2}\right)
\,q\cdot S_2\left[ 1 + {\cal O}\left(\frac{Q^2}{m_N^2}\right)\right]
\nonumber \\
&&+ \ (1\leftrightarrow 2)\label{1pi}
\\
{\vec A}^0(2) &=& (\tauvec_1 +\tauvec_2) \frac{g_A}{2\Fp^2}
\,\left[{\cal O} \left(\frac{Q^2}{m_N^3}\right)+{\cal O}\left(
\frac{Q^3}{m_N^2 M_V^2}\right)\right].
\eea
The leading part of Equation (\ref{1pi}) is nothing but the one-pion
exchange current with one-loop radiative corrections (\ref{freeA}) expressed
now in terms of a vector-dominated Dirac form factor $F_1^V$. (In fact, with
the Lagrangian (\ref{I6}), the {\it soft-pion} contribution corresponds to
(\ref{1pi}) in the limit $M_V\rightarrow \infty$.) There is no further
correction to what has already
been obtained with our Lagrangian given in its full
glory in Appendix A. This corroborates our argument
that the counter terms $\kappa_4^{(1,2)}$
cannot come from one vector-meson exchange in the limit $m_V\rightarrow
\infty$. This also establishes our assertion that vector mesons
do not modify our result on the ``chiral filter mechanism."

For completeness, we give the corresponding axial-charge
operator in coordinate space
\be
\tilde{{\cal M}}_{tree}^{\mbox{\tiny VMD}}
+\tilde{{\cal M}}_{1\pi}^{\mbox{\tiny VMD}}
= {\tilde{\cal T}}^{(1)} \,
(1+ \delta_{soft})\,\frac{1}{4\pi r}
\left[ \left(\Mp + \frac{1}{r}\right) \e^{-\Mp r} -
\left(M_V+ \frac{1}{r}\right) \e^{-M_V r} \right]\label{vector}
\ee
where $\delta_{soft} = \frac{\Mp^2}{M_V^2 - \Mp^2}$ and we have dropped the
terms of order $m_N^{-2}$. This is the vector-dominated form of $\M_{1\pi}$,
in place of (\ref{onepiloop}): The second term of
vector-meson range in (\ref{vector}) is the counterpart to the shorter-ranged
loop correction in (\ref{onepiloop}). In fermi-gas model (\ref{vector})
predicts roughly the same quenching as the loop calculation
(\ref{onepiloop}).

\newpage
\centerline{\bf FIGURE CAPTIONS}
\vskip 1cm
\noindent {\bf Figure 1}
\begin{quotation}
\noindent
Generic nuclear electroweak currents up to two body. The solid line represents
the nucleon, the blob with a cross the
coupling of electroweak fields and the shaded blob without cross
stands for the strong interactions.
\end{quotation}
\vskip 0.5cm
\noindent {\bf Figure 2}
\begin{quotation}
\noindent
Two-body exchange currents: (a) One-pion exchange; (b) two-pion exchange.
The solid blob represents a strong-interaction vertex and the shaded blob with
a cross the vertex involving an external field and strong interactions.
The solid line represents the nucleon and broken line the pion.
\end{quotation}
\vskip 0.5cm
\noindent {\bf Figure 3}
\begin{quotation}
\noindent
One-loop graphs contributing to the nucleon self-energy $\Sigma$.
As in Figure 2, the solid line represents the nucleon, the broken line the
pion.
\end{quotation}
\vskip 0.5cm
\noindent {\bf Figure 4}
\begin{quotation}
\noindent
One-loop graphs contributing to the three-point $G_\mu NN$ vertex where
$G_\mu= {\cal A}_\mu$ (${\cal V}_\mu$) is the external axial-vector (vector)
field, the encircled cross representing the field coupling.
Here and in Fig. 5, vector-field couplings are also drawn
for comparison and for later use in \cite{pmr2}.
\end{quotation}
\vskip 0.5cm
\noindent {\bf Figure 5}
\begin{quotation}
\noindent
One-loop graphs contributing to the four-point $G_\mu\pi NN$ vertex.
For axial-charge transitions, only the graphs $(a)$-$(f)$ contribute.
\end{quotation}
\vskip 0.5cm
\noindent {\bf Figure 6}
\begin{quotation}
\noindent
One-loop graphs contributing to two-body two-pion exchange currents
($(a)-(h)$), four-fermion-field contact interaction currents ($(i)-(j)$) and
``recoil" current $(k)$. The pion propagator appearing in $(a)-(j)$ is
the Feynman one while that in $(k)$ is a time-ordered one.
Only the graphs $(a)$, $(b)$, $(c)$ and $(d)$
survive for the axial-charge operator.
\end{quotation}
\vskip 0.5cm
\noindent {\bf Figure 7}
\begin{quotation}
\noindent
$4\pi r^2 \tilde{{\cal M}}_{tree}$ (solid line) and $4\pi r^2
\tilde{{\cal M}}_{loop}$ (broken
line) defined in Eqs.(\ref{msoft}) and
(\ref{mloop}) vs. $r$ in fm. Here and in Fig. 8, we have set
$\tilde{\cal T}^{(1)}=\tilde{\cal T}^{(2)}=1$.
\end{quotation}
\vskip 0.5cm
\noindent {\bf Figure 8}
\begin{quotation}
\noindent
$r[j_1 (p_F r)]^2 \tilde{{\cal M}}_{tree}$ (solid line) and $r[j_1 (p_F r)]^2
\tilde{{\cal M}}_{loop}$ (broken line) vs. $r$ with $p_F\approx 1.36$
fm$^{-1}$ (corresponding to nuclear matter density). See the caption
for Fig. 7.
\end{quotation}
\vskip 0.5cm
\noindent {\bf Figure 9}
\begin{quotation}
\noindent
The ratios of the matrix elements
$\frac{\langle {\cal M}_{X}\rangle}{\langle {\cal M}_{tree}
\rangle}$ in fermi-gas model vs. $\rho/\rho_0$ for $d= 0.5,
0.7$ fm for $X=1\pi$, $2\pi$, $1\pi+2\pi$ corresponding to one-loop correction
to the one-pion exchange graph, one-loop two-pion exchange graph
and the sum of the two, respectively.
\end{quotation}
\vskip 0.5cm
\noindent {\bf Figure 10}
\begin{quotation}
\noindent
Three-body currents: a) Genuine three-body current with Feynman pion
propagators; b) ``recoil" three-body
currents with time-ordered pion propagators; the ellipsis stands for other
time-orderings and permutations.
Both (a) and (b) are of order $O(Q^3)$ relative to the leading soft-pion term.
\end{quotation}
\vskip 0.5cm
\noindent {\bf Figure 11}
\begin{quotation}
\noindent Vector-meson contribution with the Lagrangian (\ref{I6})
to the two-body axial charge operator. $V$ and $V^\prime$ stand for vector
mesons of mass $M_V$. For the axial current, $V=\rho$ and $V^\prime=\omega$.
\end{quotation}
\end{document}